\documentclass[showkeys,superscriptaddress,floatfix,prd,10pt,aps]{revtex4-2}
\usepackage{graphicx,epstopdf}
\usepackage[caption=false]{subfig}
\usepackage{appendix}
\usepackage[T1]{fontenc}
\usepackage{lmodern}
\usepackage[dvipsnames,x11names]{xcolor}
\usepackage[colorlinks=true,linkcolor=NavyBlue,citecolor=ForestGreen,urlcolor=NavyBlue]{hyperref}
\usepackage[sort&compress]{natbib}
\usepackage{lipsum}
\usepackage{morefloats}
\usepackage[pdf]{pstricks}
\usepackage{amsmath}
\usepackage{amssymb}
\usepackage{amsfonts}
\usepackage{rotating}
\usepackage{cancel}
\usepackage{mathtools}
\usepackage{bbm}
\usepackage{dsfont}
\usepackage{bbold}
\usepackage{multirow}
\usepackage{ulem}
\usepackage{physics}
\usepackage{orcidlink}
\usepackage{colortbl}
\usepackage{dcolumn}
\usepackage{bm}

\usepackage{color,soul}

\def\pslash{p\!\!\!\slash }

\def\xslash{x\!\!\!\slash }
\def\yslash{y\!\!\!\slash }
\def\dslash{D\!\!\!\slash }

\begin{document}


\title{Gravitational form factors of $\Delta$ baryon via QCD sum rules}
\author{Z.  Dehghan\orcidlink{0000-0002-5976-9231}}
\email{zeinab.dehghan@ut.ac.ir}
\affiliation{Department of Physics, University of Tehran, North Karegar Avenue, Tehran
	14395-547, Iran}
\author{K.  Azizi\orcidlink{0000-0003-3741-2167}}
\email{kazem.azizi@ut.ac.ir, Corresponding Author}
\affiliation{Department of Physics, University of Tehran, North Karegar Avenue, Tehran
	14395-547, Iran}
\affiliation{Department of Physics, Do\v{g}u\c{s} University, Dudullu-\"{U}mraniye, 34775
	Istanbul,  T\"{u}rkiye}
\author{U.  \"{O}zdem\orcidlink{0000-0002-1907-2894}}
\email{ulasozdem@aydin.edu.tr}
\affiliation{Health Services Vocational School of Higher Education, Istanbul Aydin University, Sefakoy-Kucukcekmece, 34295 Istanbul, T\"{u}rkiye}

\begin{abstract}
The gravitational form factors of a hadron  are defined through the matrix elements of the energy-momentum tensor current, which can be decomposed into the quark and gluonic parts, between the hadronic states.
These form factors provide important information for answering fundamental questions about the distribution of the energy,  the spin,  the pressure and the shear forces inside the hadrons.    Theoretical and  experimental  studies of these form factors provide exciting insights on the inner structure and geometric shapes of hadrons.  Inspired by this, the gravitational form factors of $\Delta$ resonance are calculated by employing the  QCD sum rule approach. The acquired gravitational form factors are used to calculate the composite gravitational form factors  like the energy and angular momentum multipole form factors, D-terms related to the  mechanical properties like the internal pressure and shear forces   as well as the mass radius of the system. The predictions are  compared with the existing results in the literature.
  \end{abstract}
\date{\today}
\maketitle

\section{Introduction} 
The key subject of nonperturbative QCD is to figure out the inner structure of hadrons and their properties concerning the degrees of freedom of quarks and gluons. Different hadronic charges characterized as matrix elements of the vector, axial, and tensor currents between hadronic states contain precise information about the internal structure  distributions of different physical quantities  and geometric shapes of the hadrons. Besides the electromagnetic, axial, and tensor form factors of hadrons, the gravitational form factors (GFFs) or energy-momentum tensor form factors (EMTFFs) are also fundamental constituents to investigate the inner organizations of hadrons. These form factors give us a tool to systematically study the properties of the hadrons and to calculate several related observables such as spin, multipole form factors, mass  and mechanical radii, shear force and energy-pressure distributions inside the hadrons. Understanding the mechanical structure of hadrons is important because it gives us fundamental information about the internal structure and geometric shapes of hadrons as stated. 

In recent years, GFFs have attracted increasing interests in describing the features of hadrons with different spins because of their relation to the generalised parton distributions (GPDs).  The GPDs can be extracted from available data of hard exclusive process like deeply virtual Compton scattering (DVCS), deeply virtual meson production (DVMP), wide-angle Compton scattering (WACS), single diffractive hard exclusive processes (SDHEPs)  and  different vector-meson electro-production processes. The GFFs can be directly calculated from the  theory, as well. Comparison of  GFFs calculated from pure theory with the ones extracted from the GPDs are indirect comparison of the experimental data with theoretical predictions on many physical  observables. Such comparison for the GFFs of nucleon is made in Ref. \cite{Hashamipour:2022noy}: The consistency of the results obtained from both sides show that the mankind is in the right way regarding the  theoretical and experimental extractions of the  nucleon properties.


The GFFs for the spin-1/2 particles have been parameterized in Refs. \cite{Kobzarev:1962wt,Pagels:1966zza,Kobsarev:1970qm,Ng:1993vh}. Utilizing these parameterizations, the GFFs of spin 1/2 baryons have been studied in different phenomenological models~\cite{Polyakov:2002wz,Ji:1997gm,Schweitzer:2002nm,Jung:2013bya,Hagler:2003jd,Gockeler:2003jfa,Pasquini:2007xz,Hwang:2007tb,Abidin:2008hn,Brodsky:2008pf,Pasquini:2014vua,Chakrabarti:2015lba,Lorce:2018egm,Teryaev:2016edw,Shanahan:2018nnv, Shanahan:2018pib,Neubelt:2019sou,Anikin:2019kwi,Alharazin:2020yjv,Gegelia:2021wnj,Varma:2020crx,Fujita:2022jus,Mamo:2022eui,Pefkou:2021fni,Azizi:2019ytx,Polyakov:2018exb,Freese:2021mzg,Freese:2021qtb,Freese:2021czn,Burkert:2021ith,Burkert:2018bqq,Burkert:2023wzr,Ozdem:2020ieh,Won:2022cyy,Polyakov:2002yz,Hashamipour:2022noy,Hashamipour:2021kes}. For a spin-1 particle, the corresponding GFFs  were studied in Refs.~\cite{Polyakov:2019lbq,Cosyn:2019aio,Kim:2022wkc,Freese:2022yur,Freese:2022ibw,Freese:2019bhb,Sun:2020wfo,Epelbaum:2021ahi,Pefkou:2021fni}. The GFFs for the spin-3/2 states have also been investigated in Refs. \cite{Pefkou:2021fni,Fu:2022rkn,Alharazin:2022wjj,Perevalova:2016dln,Panteleeva:2020ejw,Cotogno:2019vjb,Kim:2020lrs,Fu:2023ijy,Fu:2023dea,Fu:2022rkn}.  
The computations of GFFs have also been extended to the $ N^*\rightarrow N $ and $N \to \Delta$ transitions in Refs.~\cite{Polyakov:2020rzq, Azizi:2020jog, Ozdem:2019pkg, Kim:2022bwn, Ozdem:2022zig}. To this end, methods like  the lattice QCD, the light-cone QCD sum rule, the chiral effective theory, the chiral quark model,  the $SU(2)$ skyrme model, the AdS/CFT correspondence and the bag model have  been used.

In the present study,  the GFFs of the  $\Delta$ baryon are calculated utilizing the three-point QCD sum rule technique, as one of the powerful and successful nonperturbative methods in hadron physics. With the help of this method, we  extract the behavior of the $\Delta$ baryon's GFFs  with respect to  $ Q^2 $  and, in connection with this, the mechanical properties of this resonance: The energy and angular momentum multipole form factors, $ \mathcal{D} $ terms related to  the internal pressure and shear forces   as well as the mass radius.
On  contrary to the electromagnetic form factors of the $\Delta$ baryon, which have been widely studied both theoretically and experimentally \cite{Fu:2022rkn, Azizi:2009egn, Kotulla:2002cg}, it is quite hard to extract  the GFFs of the $\Delta$ baryon experimentally or obtain them from the corresponding GPDs due to the short-lived nature of the  $\Delta$ baryon. 
The  GFFs of $\Delta$ baryon and the corresponding mechanical properties have theoretically been studied using relativistic covariant quark-diquark approach~\cite{Fu:2022rkn}, chiral EFT~\cite{Alharazin:2022wjj}, lattice QCD (for the gluonic part)~\cite{Pefkou:2021fni} and $SU(2)$ skyrme model based on the large $N_c$ limit \cite{Kim:2020lrs}. More systematic studies are needed to examine the features of $\Delta$ baryon GFFs. Recently, GPDs of spin-3/2 hadrons and the sum rules that connect the GPDs with the GFFs have explicitly been displayed in Refs.~\cite{Fu:2022bpf,Fu:2023dea}.

The remainder of this paper is structured as follows: In Sect. II, GFFs of $\Delta$ baryon calculated via the three-point QCD sum rules approach are introduced. The gravitational multipole form factors of $\Delta$ baryon are given in Sect. III. Numerical analysis of the GFFs and the mechanical properties of the $\Delta$ baryon are presented in Sect. IV.  In the last section, we conclude our work with a discussion of the obtained results.

\section{QCD sum rules for the gravitational form factors of the $\Delta$ baryon }\label{sec:formalism}

We use the QCD sum rules to calculate the gravitational form factors of the $\Delta $ baryon. For this purpose, we consider the following three-point correlation function,
\begin{equation}\label{eq:corrf}
\Pi_{\alpha\mu\nu\beta}(p,q) = i^2 \int d^4 x e^{-ip.x} \int d^4 y e^{ip'.y}
\langle 0 |\mathcal{T}[J_{\alpha}^{\Delta}(y)T_{\mu \nu}(0)\bar{J}_{\beta}^{\Delta}(x)]| 0 \rangle,
\end{equation}
where $\mathcal{T}$ denotes the time ordering operator, $p$ ($p'$) is the four-momentum of the initial (final) $\Delta$ baryon, $q = p - p'$ is the momentum transfer, $J_{\alpha}^{\Delta}(y)$ is  the interpolating current for the $\Delta$ state at point $y$, and $T_{\mu\nu}$ is the energy-momentum tensor (EMT) current. The interpolating current for $\Delta^+$ is given by,
\begin{equation}\label{eq:deltainterpolating}
J_{\alpha}(x)=\frac{1}{\sqrt{3}}\varepsilon^{abc}\left[
\vphantom{\int_0^{x_2}}
2 \Big(u^{aT} (x) C\gamma_{\alpha}d^{b} (x) \Big)  u^{c} (x) +
\Big(u^{aT} (x) C\gamma_{\alpha}u^{b} (x) \Big) d^{c} (x) \right],
\end{equation}
where $C$ is the charge conjugation operator; and $a, b$ and $c$ are color
indices. The EMT current has two parts: One from the quarks and another one from the gluons, as given below,
\begin{equation}
\label{eq:EMTcurrentquark1}
T_{\mu\nu} (z) =  T_{\mu\nu}^q (z) + T_{\mu\nu}^g(z),
\end{equation}
with
\begin{eqnarray}
\label{eq:EMTcurrentquark}
T_{\mu\nu}^q (z) &=& \frac{i}{2}\bigg[\bar{u}(z)\overleftrightarrow{D}_\mu \gamma_\nu u(z) 
+ \bar{u}(z)\overleftrightarrow{D}_\nu \gamma_\mu u (z)
+ \bar{d}(z)\overleftrightarrow{D}_\mu \gamma_\nu d(z) 
+\bar{d}(z)\overleftrightarrow{D}_\nu \gamma_\mu d(z) \bigg]\nonumber\\
&&
- g_{\mu\nu} \Big[\bar{u}(z)  \Big(i \overleftrightarrow{\dslash} - m_u\Big)u(z)
+\bar{d}(z)\Big( i \overleftrightarrow{\dslash} -  m_d\Big)d(z)  \Big],
\\
T_{\mu\nu}^g (z) &=& \frac{1}{4}g_{\mu\nu}G^{\rho\delta}(z)G_{\rho\delta}(z) - G_{\mu\rho}(z) G^{\rho}_{\nu}(z).\label{eq:EMTcurrentgluon}
\end{eqnarray} 
We can rewrite the second term of the quark EMT current in Eq.~\eqref{eq:EMTcurrentquark} as follows \cite{Polyakov:2018zvc},
\begin{equation}\label{eq:removable}
g_{\mu\nu} \Big[\bar{u}(z)  \Big(i \overleftrightarrow{\dslash}-m_u\Big)u(z)+\bar{d}(z)\Big( i \overleftrightarrow{\dslash}-m_d\Big)d(z) \Big] 
\simeq  
g_{\mu\nu} (1+\gamma_m)\Big(m_u \bar uu + m_d \bar dd \Big),
\end{equation}
where $\gamma_m$ denotes the anomalous dimension of the mass operator. We assume the chiral limit where $m_u = m_d = 0$. This eliminates the term in Eq.~\eqref{eq:removable}. The covariant derivative $\overleftrightarrow{D}_{\mu}$ in Eq. (\ref{eq:EMTcurrentquark}) is given by,
\begin{equation}\label{eq:deriv}
	\overleftrightarrow{D}_\mu (z) =
	\frac{1}{2} [ \overrightarrow{D}_\mu (z) - \overleftarrow{D}_\mu (z)],
\end{equation}
with,
\begin{equation}\label{eq:derivdetail}
	\overrightarrow{D}_{\mu}(z)=\overrightarrow{\partial}_{\mu}(z)-i
	\frac{g}{2}\lambda^a A^a_\mu(z), \qquad
	\overleftarrow{D}_{\mu}(z)=\overleftarrow{\partial}_{\mu}(z)+
	i\frac{g}{2}\lambda^a A^a_\mu(z),
\end{equation}
where, $\lambda^a$ are the Gell-Mann matrices and $A^a_\mu(z)$
are the external gluon fields. Using the Fock-Schwinger gauge, $z^\mu A^a_\mu(z)=0$, the gluon fields can be expressed in terms of the gluon field strength tensor by,
\begin{equation}\label{eq:gluonfield}
	A^{a}_{\mu}(z) = \int_{0}^{1}d\alpha \alpha 
	z_\xi G_{\xi\mu}^{a}(\alpha z)
	= \frac{1}{2}z_{\xi} G_{\xi\mu}^{a}(0)
	+ \frac{1}{3}z_\eta z_\xi {D}_\eta
	G_{\xi\mu}^{a}(0)+\cdots.
\end{equation}
To calculate the derivative terms of the quark part of the EMT current, we evaluate the EMT current at point $z$ in Eq.~\eqref{eq:corrf} and finally take the limit $z \to 0$. In this limit, Eq.~\eqref{eq:gluonfield} shows that the gluon field vanishes and therefore the covariant derivatives in Eq.~\eqref{eq:derivdetail} become partial derivatives.

In the QCD sum rule approach, we define the correlation function in two different representations: One based on the hadronic degrees of freedom and is called the physical (phenomenological) side and the other  based on QCD degrees of freedom and is called the QCD (theoretical) side. The double Borel transformations with respect to the momentum squared of the initial and final states are applied to both sides  to remove/suppress the contributions coming from the subtraction terms/higher states and continuum. A continuum subtraction procedure supplied by quark-hadron duality assumption is also applied to further suppress the contributions of the higher states and enhance the ground state contribution. The form factors are obtained by matching the coefficients of the same Lorentz structures of both representations.

\subsection{Physical side of the correlation  function}\label{subsec:physical side}

We start by evaluating the correlation function in Eq. (\ref{eq:corrf}) using hadronic parameters. For this purpose, we insert two complete sets of the intermediate states $\Delta(p',s')$ and $\Delta(p,s)$ into Eq.~\eqref{eq:corrf} and perform the four-integrals over $x$ and $y$, which ends up in
\begin{equation}\label{eq:physicalcorr}
\Pi_{\alpha\mu\nu\beta}^\text{Had}(p,q) = 
\sum_{s{'}}\sum_{s} \frac{\langle0|J_{\alpha}^{\Delta}|{\Delta(p',s')}\rangle\langle {\Delta(p',s')}
|T_{\mu \nu}(0)|
{\Delta(p,s)}\rangle\langle {\Delta(p,s)}
|\bar{J}_{\beta}^{\Delta}| 0 \rangle}
{(m^2-p'^2)(m^2-p^2)} 
+\cdots,
\end{equation}
where $m = m_{\Delta}$ and the dots indicate the higher states and continuum contributions. The matrix element of the EMT current between $\Delta$ states can be expressed in terms of ten form factors~\cite{Cotogno:2019vjb, Kim:2020lrs}
\begin{align}\label{eq:matrix element}
\begin{aligned}
\langle {\Delta(p',s')}
|T_{\mu \nu}(0)|
{\Delta(p,s)}\rangle &= 
- \bar{u}_{\alpha'}(p',s')\Big\{
\frac{P_{\mu}P_{\nu}}{m}
\Big(g^{\alpha' \beta'} F_{1,0}(Q^2) 
- \frac{\Delta^{\alpha'}\Delta^{\beta'}}{2 m^2} 
F_{1,1}(Q^2)\Big)\\
&\hspace{-3 cm}+
\frac{(\Delta_{\mu}\Delta_{\nu} - g_{\mu\nu} \Delta^2)}{4 m}
\Big(g^{\alpha' \beta'} F_{2,0}(Q^2) 
- \frac{\Delta^{\alpha'}\Delta^{\beta'}}{2 m^2} 
F_{2,1}(Q^2)\Big)
+ m g_{\mu\nu}
\Big(g^{\alpha' \beta'} F_{3,0}(Q^2) 
- \frac{\Delta^{\alpha'}\Delta^{\beta'}}{2 m^2} 
F_{3,1}(Q^2)\Big)\\
&\hspace{-3 cm}+
\frac{i}{2}
\frac{(P_{\mu} \sigma_{\nu\rho} + P_{\nu} \sigma_{\mu\rho})
\Delta^{\rho}}{m}
\Big(g^{\alpha' \beta'} F_{4,0}(Q^2) 
- \frac{\Delta^{\alpha'}\Delta^{\beta'}}{2 m^2} 
F_{4,1}(Q^2)\Big)
- \frac{1}{m}\Big(
g^{\alpha'}_{\mu}\Delta_{\nu}\Delta^{\beta'} 
+ g^{\alpha'}_{\nu}\Delta_{\mu}\Delta^{\beta'}
+ g^{\beta'}_{\mu}\Delta_{\nu}\Delta^{\alpha'}
\\
&\hspace{-3 cm}
+ g^{\beta'}_{\nu}\Delta_{\mu}\Delta^{\alpha'}
-2 g_{\mu\nu} \Delta^{\alpha'}\Delta^{\beta'} 
- \Delta^2 g^{\alpha'}_{\mu} g^{\beta'}_{\nu}
- \Delta^2 g^{\alpha'}_{\nu} g^{\beta'}_{\mu}
\Big)F_{5,0}(Q^2)
+
m \Big(
g^{\alpha'}_{\mu} g^{\beta'}_{\nu}
+ g^{\alpha'}_{\nu} g^{\beta'}_{\mu}
\Big)F_{6,0}(Q^2)
\Big\} u_{\beta'}(p,s), 
\end{aligned}
\end{align}
where $u_{\beta'}(p,s)$ is the Rarita-Schwinger spinor with momentum $p$ and spin $s$, $P = (p + p')/2$, $\Delta = p' - p$, $Q^2 = - \Delta^2$ and $ F_{i,k} $ are GFFs. We consider the full system that includes both the quark and gluon contributions to the EMT, given by Eqs.~\eqref{eq:EMTcurrentquark} and \eqref{eq:EMTcurrentgluon}, implying the conservation of the total current. Therefore, the non-conserved form factors $F_{i,k} (i=3,6)$ vanish, while the conserved ones $F_{i,k} (i=1, 2, 4, 5)$ remain alive. By using the residue of the $\Delta$ baryon ($\lambda_{\Delta}$), one can define the following matrix element,
\begin{equation}\label{eq:residue}
\langle0|J_{\alpha}^{\Delta}|{\Delta(p',s')}\rangle = \lambda_{\Delta} u_{\alpha}(p',s').
\end{equation}
We introduce the spin summation of the Rarita-Schwinger spinor for the $\Delta$ baryon as below,
\begin{equation}\label{eq:sspin}
\sum_{s'}{u_\alpha}(p',s') \bar{u}_{\alpha'}(p',s') = 
- ({\pslash'} + m)
\Big[g_{\alpha\alpha'} - \frac{\gamma_{\alpha}\gamma_{\alpha'}}{3}
-\frac{2 p'_{\alpha} p'_{\alpha'}}{3 m^2}
+\frac{p'_{\alpha} \gamma_{\alpha'} - p'_{\alpha'} \gamma_{\alpha}}{3 m}
\Big].
\end{equation}
Using Eqs.~\eqref{eq:matrix element}, \eqref{eq:residue} and \eqref{eq:sspin} in Eq.~\eqref{eq:physicalcorr}, we derive the following  expression for the $\Delta \to \Delta$ transition three-point correlation function,
\begin{align}\label{eq:physicalcorrfun}
\begin{aligned}
\Pi_{\alpha\mu\nu\beta}^\text{Had}&(p,q) = 
\frac{-\lambda_{\Delta}^2}{(m^2-p'^2)(m^2-p^2)}
({\pslash'} + m)
\Big[g_{\alpha\alpha'} - \frac{\gamma_{\alpha}\gamma_{\alpha'}}{3}
-\frac{2 p'_{\alpha} p'_{\alpha'}}{3 m^2}
+\frac{p'_{\alpha} \gamma_{\alpha'} - p'_{\alpha'} \gamma_{\alpha}}{3 m}
\Big]
\\
&\times\Big\{
\frac{P_{\mu}P_{\nu}}{m}
\Big(g^{\alpha' \beta'} F_{1,0}(Q^2) 
- \frac{\Delta^{\alpha'}\Delta^{\beta'}}{2 m^2} 
F_{1,1}(Q^2)\Big)
+
\frac{(\Delta_{\mu}\Delta_{\nu} - g_{\mu\nu} \Delta^2)}{4 m}
\Big(g^{\alpha' \beta'} F_{2,0}(Q^2) 
- \frac{\Delta^{\alpha'}\Delta^{\beta'}}{2 m^2} 
F_{2,1}(Q^2)\Big)
\\
&+
\frac{i}{2}
\frac{(P_{\mu} \sigma_{\nu\rho} + P_{\nu} \sigma_{\mu\rho})
\Delta^{\rho}}{m}
\Big(g^{\alpha' \beta'} F_{4,0}(Q^2) 
- \frac{\Delta^{\alpha'}\Delta^{\beta'}}{2 m^2} 
F_{4,1}(Q^2)\Big)
- \frac{1}{m}\Big(
g^{\alpha'}_{\mu}\Delta_{\nu}\Delta^{\beta'} 
+ g^{\alpha'}_{\nu}\Delta_{\mu}\Delta^{\beta'}
\\
&+
g^{\beta'}_{\mu}\Delta_{\nu}\Delta^{\alpha'}
+g^{\beta'}_{\nu}\Delta_{\mu}\Delta^{\alpha'}
-2 g_{\mu\nu} \Delta^{\alpha'}\Delta^{\beta'} 
- \Delta^2 g^{\alpha'}_{\mu} g^{\beta'}_{\nu}
- \Delta^2 g^{\alpha'}_{\nu} g^{\beta'}_{\mu}
\Big)F_{5,0}(Q^2)\Big\}\\
&\times
({\pslash} + m)
\Big[g_{\beta'\beta} - \frac{\gamma_{\beta'}\gamma_{\beta}}{3}
-\frac{2 p_{\beta'} p_{\beta}}{3 m^2}
+\frac{p_{\beta'} \gamma_{\beta} - p_{\beta} \gamma_{\beta'}}{3 m}
\Big]+\cdots.  
\end{aligned}
\end{align}

In principle, the physical side of the correlation function can be obtained using the above equation. However, at this point we face with two problems that prevent the calculations  being reliable: All Lorentz structures are not independent and the correlation function can also receive contributions from spin-1/2 particles, which should be eliminated.  Indeed, the matrix element of the current $J_{\alpha}$ between vacuum and spin-1/2 baryons is nonzero and is determined as
\begin{equation}\label{spin12}
\langle0\mid J_{\alpha}(0)\mid B(p,s=1/2)\rangle=(A  p_{\alpha}+B\gamma_{\alpha})u(p,s=1/2).
\end{equation}
As is seen the unwanted spin-1/2 contributions are proportional to $\gamma_\alpha$ and $p_\alpha$.  By multiplying both sides with $\gamma^\alpha$ and employing the condition $\gamma^\alpha J_\alpha = 0$ one can specify the constant A in terms of B. To eliminate the spin-1/2 contributions and acquire only independent structures in the correlation function, we use the  ordering for Dirac matrices as $\gamma_{\alpha}\pslash'\pslash\gamma_{\mu} \gamma_{\nu}\gamma_{\beta}$ and remove terms  with $\gamma_\alpha$ at the beginning, $\gamma_\beta$ at the end and those proportional to $p'_\alpha$ and 
$p_\beta$.  
After all manipulations mentioned above done, we get the final form of the physical side of the correlation function as follows:
\begin{align}\label{eq:HadStr}
\Pi_{\alpha\mu\nu\beta}^\text{Had}(Q^2) &= \lambda_{\Delta}^2 e^{-\frac{m^2}{M^2}} \Big[ 
\Pi_{1}^\text{Had} (Q^2) p_{\alpha}p_{\mu}p_{\nu}p'_{\beta} \pslash
+ \Pi_{2}^\text{Had} (Q^2) p_{\alpha}p_{\mu}p'_{\nu}p'_{\beta} \pslash
+ \Pi_{3}^\text{Had} (Q^2) p_{\alpha}p'_{\mu}p'_{\nu}p'_{\beta} \pslash
+ \Pi_{4}^\text{Had} (Q^2) p_{\mu}p_{\nu} g_{\alpha\beta} \pslash \nonumber \\
&+ \Pi_{5}^\text{Had} (Q^2) p_{\mu}p'_{\nu} g_{\alpha\beta} \pslash 
+ \Pi_{6}^\text{Had} (Q^2) p'_{\mu}p'_{\nu} g_{\alpha\beta} \pslash
+ \Pi_{7}^\text{Had} (Q^2) p'_{\beta}p'_{\nu} g_{\alpha\mu} \pslash +\cdots\Big],
\end{align}
where the double Borel transformation on the variables $p^2$ and $p'^2$ with Borel parameter $M^2$ is applied. The initial and final states of the process involve $\Delta$ baryons, which have the same Borel mass parameter $M_i^2 = M_f^2 = 2 M^2$.
The functions $\Pi_{i}^\text{Had} (Q^2)$ are functions of  the GFFs and other hadronic parameters. We kept only the Lorentz structures that we use to calculate the conserved GFFs and moved the others inside the dots.

\subsection{QCD side of the correlation function}\label{subsec:QCD side}

Having the expression of the correlation function from the physical side, let us turn our attention to the evaluation of correlation function from QCD side. To this end, we need to insert the explicit forms of the  EMT current and interpolating current of the $\Delta$ baryon into the correlation function.  Substituting  $\Delta$'s interpolating current and the EMT current of Eqs.~\eqref{eq:EMTcurrentquark} and \eqref{eq:EMTcurrentgluon} into the three-point correlation function of Eq.~\eqref{eq:corrf}, we get,
\begin{equation}\label{eq:contractions}
	\Pi_{\alpha\mu\nu\beta}^\text{QCD}(p,q) = \dfrac{i^2}{6} \varepsilon^{abc} \varepsilon^{a'b'c'} 
	\int d^4 x e^{-ip.x} \int d^4 y e^{ip'.y}
	\Big( \Gamma^q_{\alpha\mu\nu\beta} + \Gamma^g_{\alpha\mu\nu\beta} \Big).
\end{equation}
The quark and gluon contributions of the EMT current yield $\Gamma^q$ and $\Gamma^g$, respectively. Using Wick's theorem, $\Gamma^q$ and $\Gamma^g$ are obtained in terms of the quark propagators. The expressions for $\Gamma^q$ and $\Gamma^g$ are too long to show here, so we refer to Eqs.~\eqref{eq:qcontractions} and \eqref{eq:gcontractions} in the Appendix~\ref{Calcu}.   

Substituting the light quark propagators in Eqs.~\eqref{eq:qcontractions} and \eqref{eq:gcontractions} and employing covariant derivatives of Eq.~\eqref{eq:deriv} and then considering $z\rightarrow 0$, we get, 
\begin{equation}\label{eq:qcd}
	\Pi_{\alpha\mu\nu\beta}^{\text{QCD}}(p,q) = \int d^4 x e^{-ip.x} \int d^4 y e^{ip'.y} \Gamma_{\alpha\mu\nu\beta} (x,y),
\end{equation}
with
\begin{equation}\label{eq:totalGamma}
	\Gamma_{\alpha\mu\nu\beta} (x,y) = \Big\{ \Gamma_{\alpha\mu\nu\beta}^{(P)}
	+ \Gamma_{\alpha\mu\nu\beta}^{(3D)}
	+ \Gamma_{\alpha\mu\nu\beta}^{(4D,q)}
	+ \Gamma_{\alpha\mu\nu\beta}^{(5D)} 
	+ \mu \leftrightarrow \nu \Big\} 
	+ \Gamma_{\alpha\mu\nu\beta}^{(4D,g)}.
\end{equation}
where the correlation function has a perturbative part $\Gamma^{(P)}$ and non-perturbative parts $\Gamma^{(3D)}$, $\Gamma^{(4D)}$ and $\Gamma^{(5D)}$ in three, four and five dimensions, respectively, which are shown in the Appendix~\ref{Calcu}. The four-dimensional non-perturbative parts for quark and gluon, $\Gamma^{(4D,q)}$ and $\Gamma^{(4D,g)}$, involve the products of two gluon field strength tensors $G^{A}_{\alpha\beta}$, which lead to gluon condensation as explained in the Appendix~\ref{gluoncondens}.

We transform the calculations to momentum space, by  employing ~\cite{Azizi:2017ubq},
\begin{equation}\label{eq:DdimensionalSchwingerParameterization}
\dfrac{1}{(R^2)^{n_j}} = \int \frac{d^D k_j}{(2 \pi)^D} e^{-ik_j.R} 
\,i (-1)^{n_j + 1} 2^{D - 2 n_j} \pi^{D/2} 
\dfrac{\Gamma[D/2 - n_j]}{\Gamma[n_j]} {\Big(\frac{-1}{k_j^2}\Big)}^{D/2 - n_j},
\end{equation}
where $R = x, y$ or $y-x$ and we set $x_{\mu} = i \partial/ {\partial p_{\mu}}$ and $y_{\mu} = -i \partial/ {\partial p'_{\mu}}$. The integrals over $x$ and $y$ in D dimensions produce two Dirac Delta functions and simplify two of  the D-dimensional integrals over $k_j$. The final integral takes simple forms after Feynman parameterizations. To perform them,  we  apply  the general formula presented in ~\cite{Azizi:2017ubq}, which takes the following form in the simplest case:
\begin{equation}\label{eq:Dint}
\int d^D\ell \frac{1}{(\ell^2 + L)^n} = 
\dfrac{i \pi^{D/2} (-1)^n \Gamma[n-D/2]}{\Gamma[n] (-L)^{n-D/2}}.
\end{equation}
Following these calculations, the QCD side of the correlation function is derived as the double dispersion integrals shown below,
\begin{equation}\label{eq:spectral density}
\Pi_{i}^{\text{QCD}}(Q^2) = \int_{0}^{s_0} ds \int_{0}^{s_0} ds' 
\frac{\rho_i(s,s',Q^2)}{(s-p^2)(s'-p'^2)},
\end{equation}
where $s_0$ is the continuum. The imaginary parts of the $\Pi_{i}^{\text{QCD}}(Q^2)$ define the spectral densities $\rho_i(s,s',Q^2)$, such that $\rho_i(s,s',Q^2) = Im[\Pi_{i}^{\text{QCD}}(Q^2)]/\pi$. To determine the imaginary parts of different structures, we use,
\begin{equation}\label{eq:imarinarypart}
\Gamma[D/2 - n] {\Big(\frac{-1}{L}\Big)}^{D/2 - n} = 
\frac{(-1)^{n-1}}{(n-2)!} (-L)^{n-2} \ln [-L].
\end{equation}
The expressions for the spectral densities $\rho_i(s,s',Q^2)$ are very lengthy  and, for the sake of simplicity, we do not present them explicitly.

In parallel to physical side, we consider the same ordering  for Dirac matrices and procedure for the elimination of  the spin-$1/2$ pollution.  We apply the  double Borel transformation to the QCD side  and obtain,
\begin{align}\label{eq:QCDStr}
\Pi_{\alpha\mu\nu\beta}^\text{QCD}(Q^2) &= \int_{0}^{s_0} ds \int_{0}^{s_0} ds' 
e^{-s/{2M^2}} e^{-s'/{2M^2}}
\Big[ 
\Pi_{1}^\text{QCD} (Q^2, s, s') p_{\alpha}p_{\mu}p_{\nu}p'_{\beta} \pslash
+ \Pi_{2}^\text{QCD} (Q^2, s, s') p_{\alpha}p_{\mu}p'_{\nu}p'_{\beta} \pslash \nonumber\\
&+ \Pi_{3}^\text{QCD} (Q^2, s, s') p_{\alpha}p'_{\mu}p'_{\nu}p'_{\beta} \pslash
+ \Pi_{4}^\text{QCD} (Q^2, s, s') p_{\mu}p_{\nu} g_{\alpha\beta} \pslash 
+ \Pi_{5}^\text{QCD} (Q^2, s, s') p_{\mu}p'_{\nu} g_{\alpha\beta} \pslash \nonumber\\
&+ \Pi_{6}^\text{QCD} (Q^2, s, s') p'_{\mu}p'_{\nu} g_{\alpha\beta} \pslash
+ \Pi_{7}^\text{QCD} (Q^2, s, s') p'_{\beta}p'_{\nu} g_{\alpha\mu} \pslash + \cdots\Big],
\end{align}
in terms of the selected structures.
Matching the same structures from the QCD and physical sides , the GFFs for the $\Delta$ baryon are derived. We again do not show the obtained sum rules in this step. 
\section{Gravitational Multipole Form Factors}\label{sec:GMFFs}

Having determined the seven conserved GFFs for the $\Delta \to \Delta$ graviton-like transition we can now define some composite observables in terms of GFFs. Such observables provide with us useful information about the inner structure, distributions of different charges and geometric shape of the hadron under consideration. Future experiments may provide opportunity for such observales to be measured. Hence, we provide sum inputs to be compared with possible related future experimental data. To this end, we use the following definitions for the kinematical variables $P^{\mu}$, $\Delta^{\mu}$ and momentum transfer squared $Q^2$ in the Breit frame,
\begin{equation}\label{eq:Breit}
P^{\mu} = (E, \vec{0}), \qquad \Delta^{\mu} = (0, \vec{\Delta}),
\qquad Q^2 = - \Delta^2 = 4 (E^2 - m^2).
\end{equation}
In this frame, we can express the gravitational multipole form factors (GMFFs) of the $\Delta$ baryon in terms of the conserved GFFs as follows ~\cite{Kim:2020lrs},
\begin{eqnarray}
\label{eq:GMFFs}
\varepsilon_0(Q^2) &=& F_{1,0}(Q^2)  
- \frac{Q^2}{6 m^2} \bigg[-\frac{5}{2} F_{1,0}(Q^2) - F_{1,1}(Q^2) 
-\frac{3}{2} F_{2,0}(Q^2) + 4 F_{5,0}(Q^2) + 3 F_{4,0}(Q^2)
\bigg] \nonumber\\
&&+ \frac{(Q^2)^2}{12 m^4} \bigg[\frac{1}{2} F_{1,0}(Q^2) 
+ F_{1,1}(Q^2) + \frac{1}{2} F_{2,0}(Q^2) + \frac{1}{2} F_{2,1}(Q^2)
- 4 F_{5,0}(Q^2) - F_{4,0}(Q^2) - F_{4,1}(Q^2)
\bigg]
\nonumber\\
&&+ \frac{(Q^2)^3}{48 m^6} \bigg[-\frac{1}{2} F_{1,1}(Q^2) 
- \frac{1}{2} F_{2,1}(Q^2) + F_{4,1}(Q^2)
\bigg],
\\
\varepsilon_2(Q^2) &=& -\frac{1}{6} \bigg[ 
F_{1,0}(Q^2) + F_{1,1}(Q^2) -4 F_{5,0}(Q^2)
\bigg] \nonumber\\
&&+ \frac{Q^2}{12 m^2} \bigg[\frac{1}{2} F_{1,0}(Q^2) + F_{1,1}(Q^2) 
+\frac{1}{2} F_{2,0}(Q^2) +\frac{1}{2} F_{2,1}(Q^2) - 4 F_{5,0}(Q^2) - F_{4,0}(Q^2)
- F_{4,1}(Q^2)
\bigg] \nonumber\\
&&+ \frac{(Q^2)^2}{48 m^4} \bigg[-\frac{1}{2} F_{1,1}(Q^2) 
- \frac{1}{2} F_{2,1}(Q^2) + F_{4,1}(Q^2)
\bigg],
\\
\mathcal{J}_1(Q^2) &=& \frac{1}{3} F_{4,0}(Q^2) 
- \frac{Q^2}{15 m^2} \bigg[F_{4,0}(Q^2) + F_{4,1}(Q^2) 
+ 5 F_{5,0}(Q^2)
\bigg] + \frac{(Q^2)^2}{60 m^4} F_{4,1}(Q^2) ,
\\
\mathcal{J}_3(Q^2) &=& - \frac{1}{6} \bigg[F_{4,0}(Q^2) + F_{4,1}(Q^2)\bigg]
+ \frac{Q^2}{24 m^2} F_{4,1}(Q^2) ,
\label{eq:GMFFsJ3}
\\
D_0(Q^2) &=& F_{2,0}(Q^2) - \frac{16}{3} F_{5,0}(Q^2) 
- \frac{Q^2}{6 m^2} \bigg[F_{2,0}(Q^2) + F_{2,1}(Q^2) 
- 4 F_{5,0}(Q^2)
\bigg] + \frac{(Q^2)^2}{24 m^4} F_{2,1}(Q^2),
\\
D_2(Q^2) &=& \frac{4}{3} F_{5,0}(Q^2),
\\
D_3(Q^2) &=& \frac{1}{6} \bigg[-F_{2,0}(Q^2) - F_{2,1}(Q^2) 
+ 4 F_{5,0}(Q^2)
\bigg] + \frac{Q^2}{24 m^2} F_{2,1}(Q^2) ,
\end{eqnarray} 
where  $\varepsilon_0 (Q^2)$, $\varepsilon_2 (Q^2)$, $\mathcal{J}_1 (Q^2)$ and $\mathcal{J}_3 (Q^2)$ are energy-monopole, energy-quadrupole, angular momentum-dipole and angular momentum-octupole form factors, respectively. The form factors $D_{0, 2, 3} (Q^2)$ are related to the internal pressures and shear forces ~\cite{Polyakov:2018zvc}. These form factors are used to define the generalized D-terms $\mathcal{D}_{0,2,3}$ of $\Delta$ baryon in the following way \cite{Panteleeva:2020ejw},
\begin{eqnarray}
\label{eq:Dterms}
\mathcal{D}_0 &=& D_0(0), \nonumber \\
\mathcal{D}_2 &=& D_2(0) + \frac{2}{m^2}\int^{\infty}_{0} dQ^2 D_3(Q^2), \nonumber \\
\mathcal{D}_3 &=& - \frac{5}{m^2}\int^{\infty}_{0} dQ^2 D_3(Q^2) .
\end{eqnarray}
The generalized D-terms are dimensionless quantities that characterize the elastic properties of hadrons. The mean square radius of the energy density, also known as the mass radius, is another important mechanical property of $\Delta$ baryon. It is given by the following formula ~\cite{Polyakov:2018zvc, Kim:2020lrs},
\begin{equation}\label{eq:radius}
\langle r^2_E\rangle = 6 \frac{d \varepsilon_0(k)}{dk} |_{k=0}
\end{equation}
In the following section, we will perform numerical analysis of the obtained GFFs and other observables made of these GFFs and discuss their values at zero momentum transfer.

\section{Numerical results}\label{sec:results}

In this section, we numerically analyze the form factors derived from the sum rules in the previous sections. 
The values of some input parameters are given as: $m_u=m_d=0$, $m_{\Delta} = 1.23~\mathrm{GeV}$,  
$\lambda_{\Delta} = 0.038~\mathrm{GeV}^3$ \cite{Aliev:2007pi},  
$\langle \bar{q}q \rangle (1\mbox{GeV}) =(-0.24\pm 0.01)^3$ $\mathrm{GeV}^3$ \cite{Belyaev:1982sa},  
$m_{0}^2 = (0.8 \pm 0.1)$ $\mathrm{GeV}^2$ \cite{Belyaev:1982sa}, 
$\langle \frac{\alpha_s}{\pi} G^2 \rangle   = (0.012\pm0.004)$ $~\mathrm{GeV}^4 $\cite{Belyaev:1982cd},  and 
$\alpha_s = (0.118 \pm 0.005)$ \cite{DELPHI:1993ukk}.  
In addition to these input parameters, there are two more auxiliary parameters called the Borel parameter $M^2$ and the continuum threshold $s_0$ that we use for the sum rules.
According to the philosophy of the QCD sum rules, these auxiliary parameters should not affect the physical quantities.  
However, in practice, it is not possible to provide such a situation.
Therefore, we look for working regions where the GFFs have weak dependence on these helping parameters. The residual dependencies appear as the uncertainties in the final results. The continuum threshold $s_0$ is associated with the energy of the first exited state. To restrict the Borel parameter, we require the pole dominance and convergence of the operator product expansion (OPE): The perturbative contribution exceeds the total nonperturbative one and the higher the dimension of the nonperturbative operator the lower its contribution. Our calculations reveal the following working regions for the $s_0$ and $M^2$,
\begin{align}\label{eq:working region}
2.9 ~\text{GeV}^2 \leqslant s_0 \leqslant 3.3 ~\text{GeV}^2, \nonumber\\
2.0 ~\text{GeV}^2 \leqslant M^2 \leqslant 3.0 ~\text{GeV}^2.
\end{align}

\begin{figure}[!htb]
\centering
\includegraphics[width=0.35\textwidth]{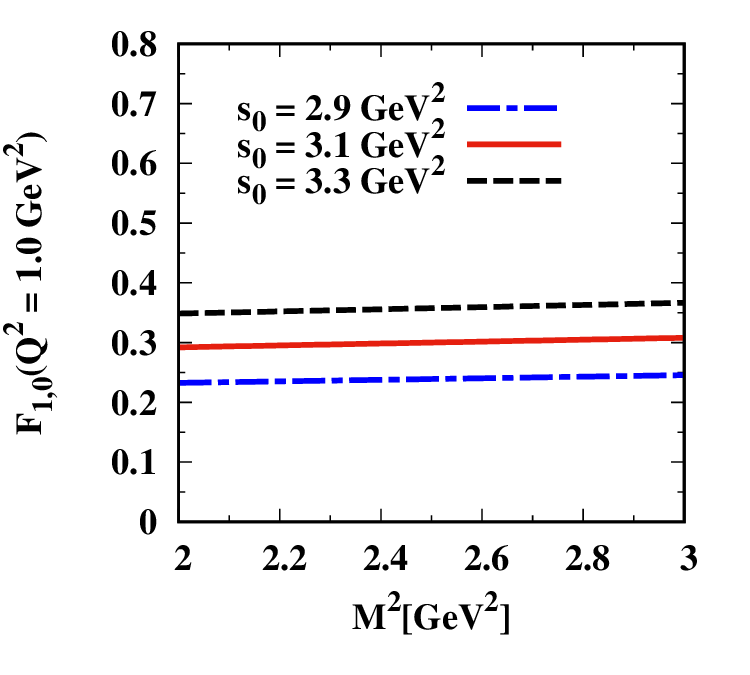}~~~~~~~~
\includegraphics[width=0.35\textwidth]{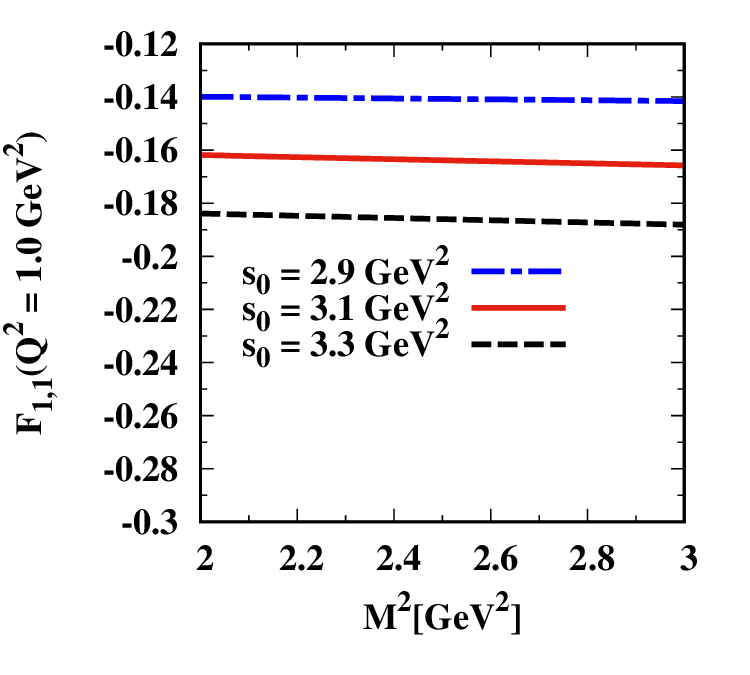}\\
\vspace{-0.1cm}
\includegraphics[width=0.35\textwidth]{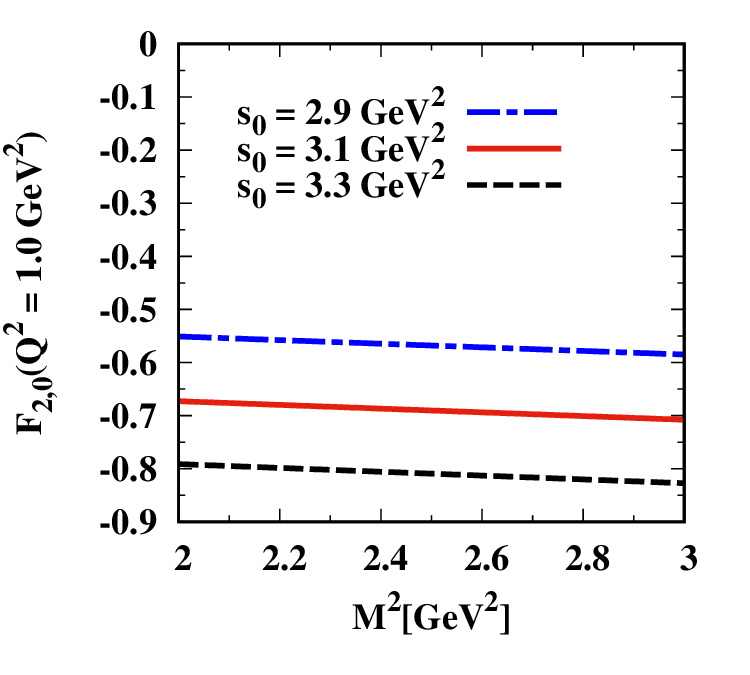}~~~~~~~~
\includegraphics[width=0.35\textwidth]{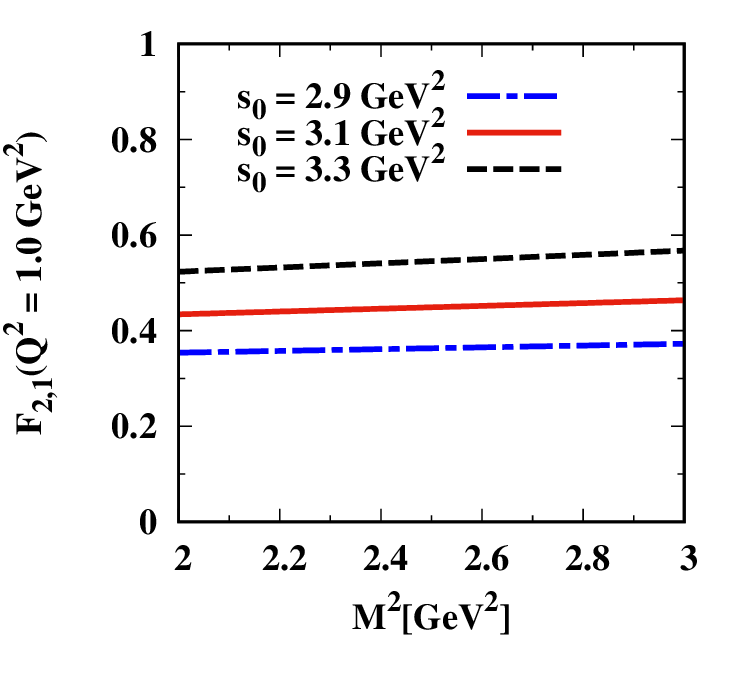}\\
\vspace{-0.1cm}
\includegraphics[width=0.35\textwidth]{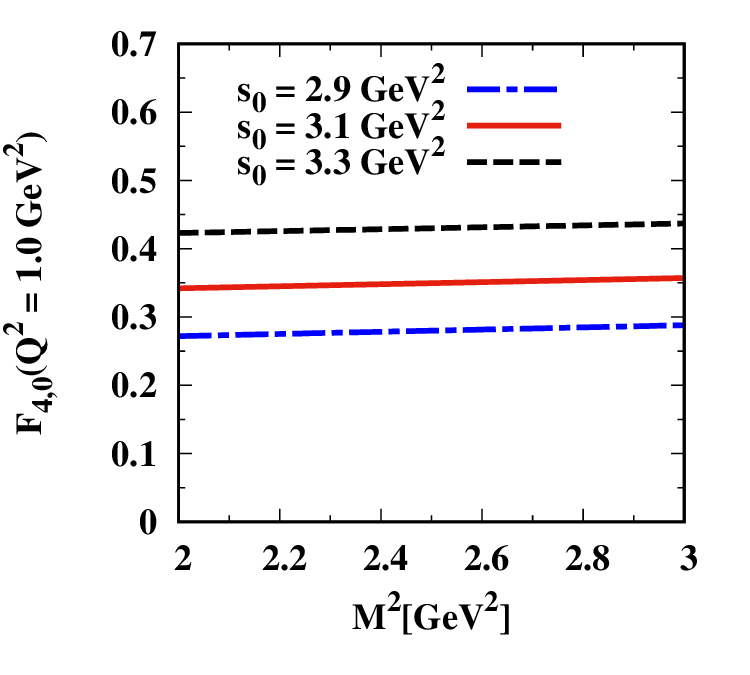}~~~~~~~~
\includegraphics[width=0.35\textwidth]{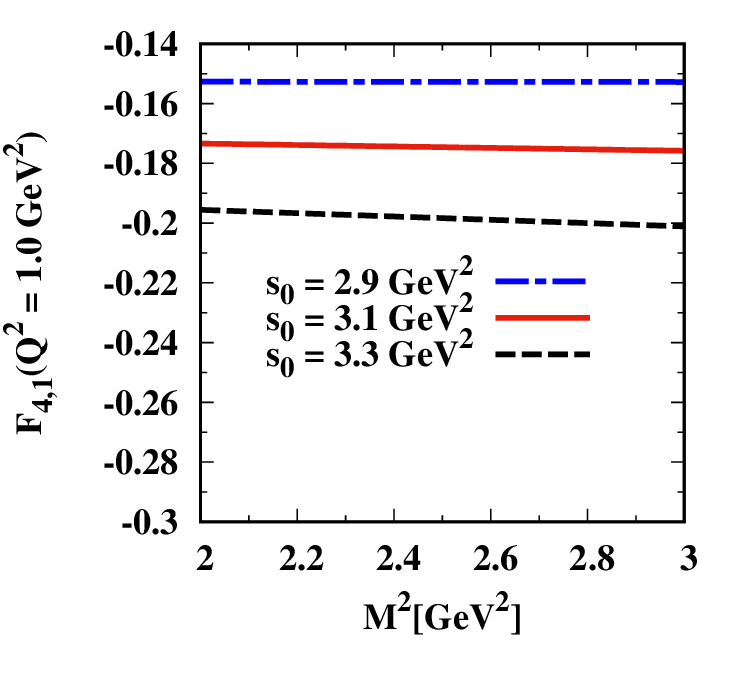}\\
\vspace{-0.1cm}
\includegraphics[width=0.35\textwidth]{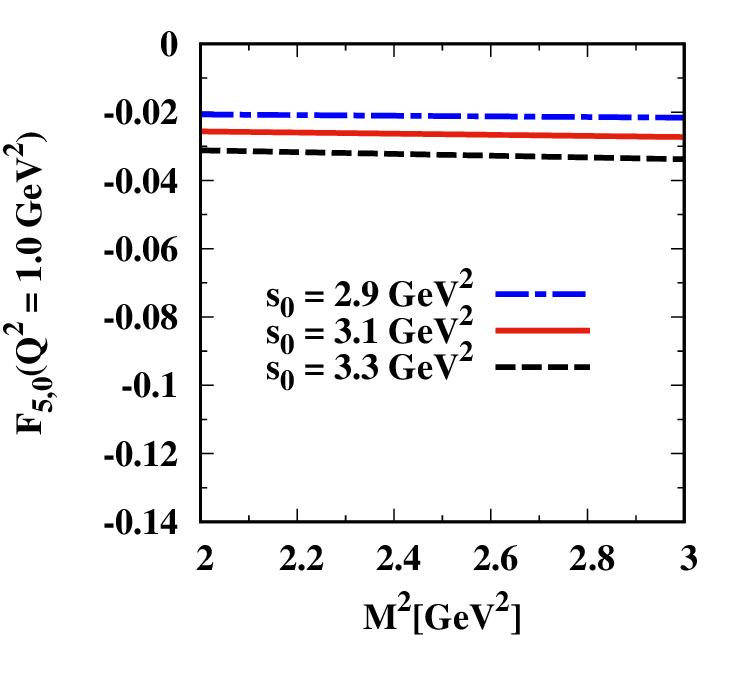}~~~~~~~~
\caption{The dependence of the GFFs of $\Delta$ on $M^2$ at $Q^2$ = 1.0~GeV$^2$ for three values of the continuum threshold $s_0$.}
\label{fig:Msqfigs}
\end{figure}

\begin{figure}[!htb]	
	\centering
	\includegraphics[width=0.35\textwidth]{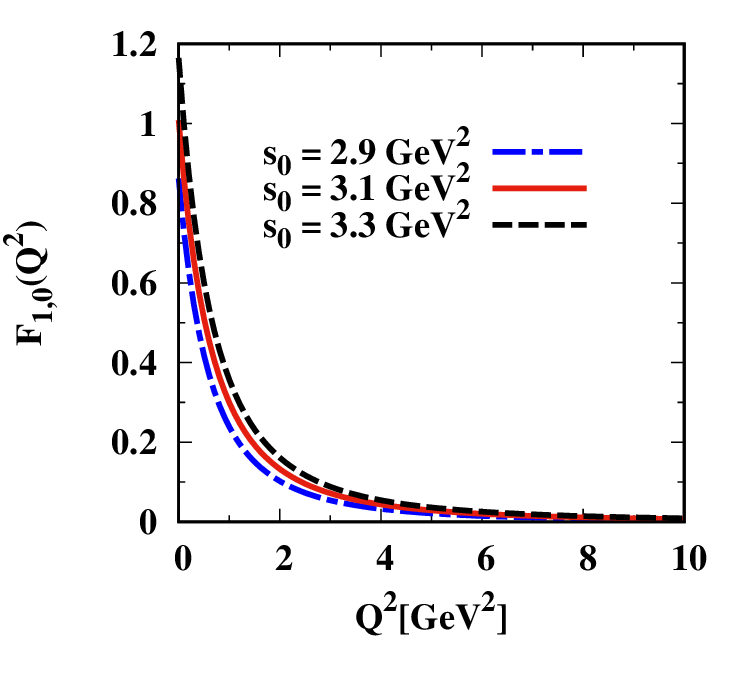}~~~~~~~~
	\includegraphics[width=0.35\textwidth]{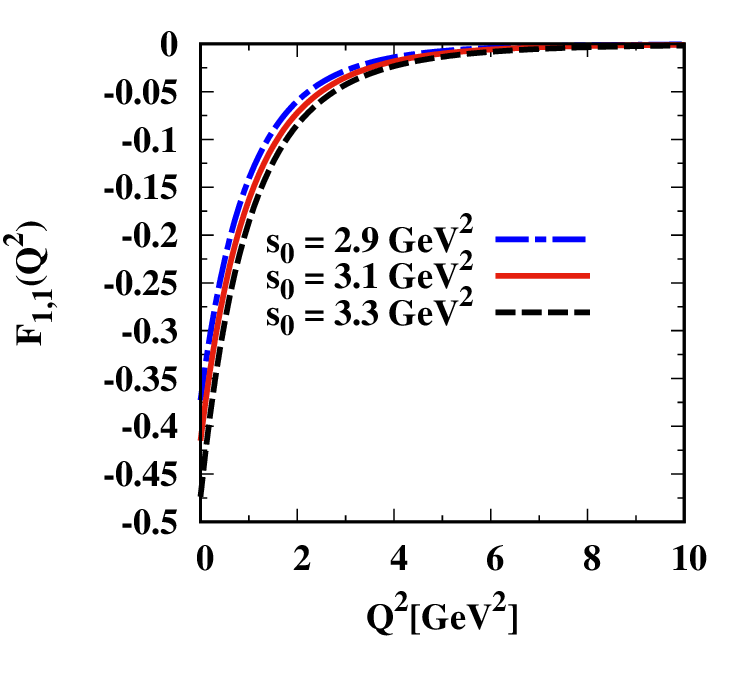}\\
	\vspace{-0.1cm}
	\includegraphics[width=0.35\textwidth]{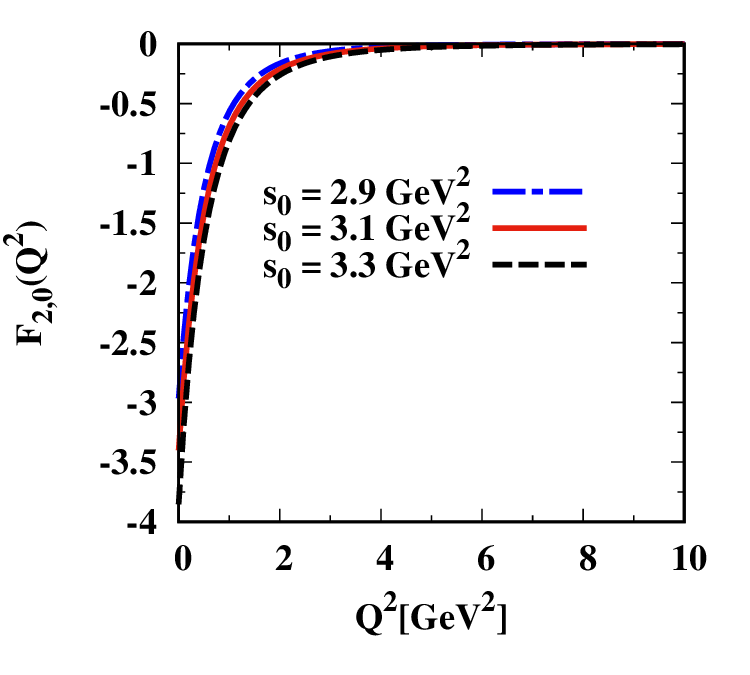}~~~~~~~~
	\includegraphics[width=0.35\textwidth]{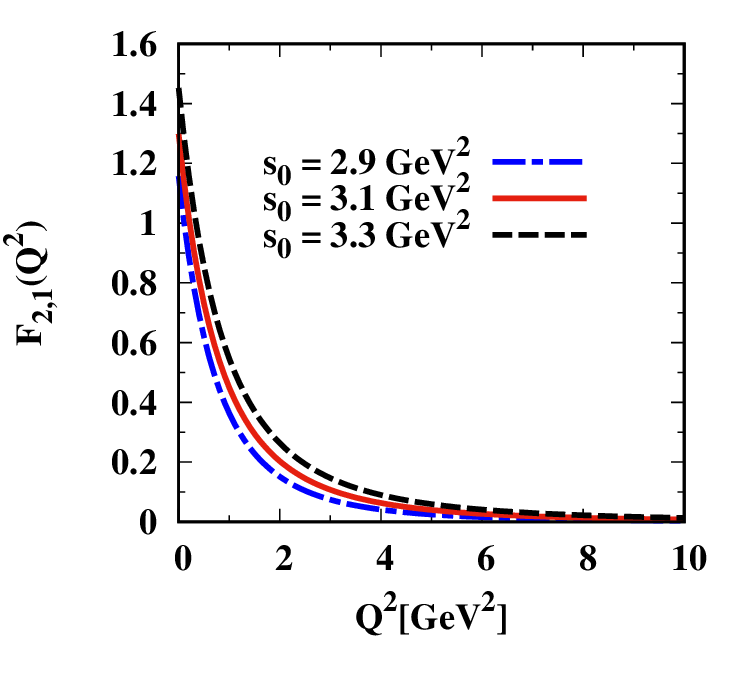}\\
	\vspace{-0.1cm}
	\includegraphics[width=0.35\textwidth]{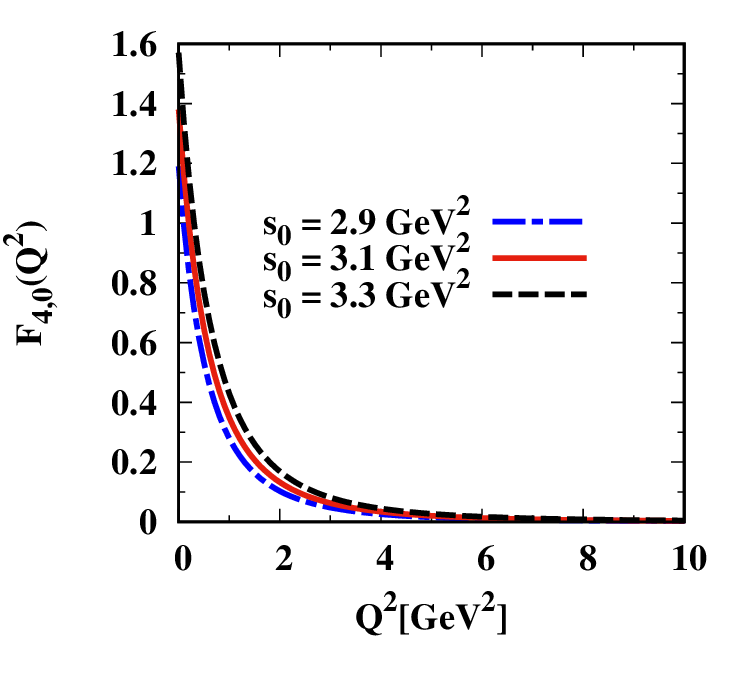}~~~~~~~~
	\includegraphics[width=0.35\textwidth]{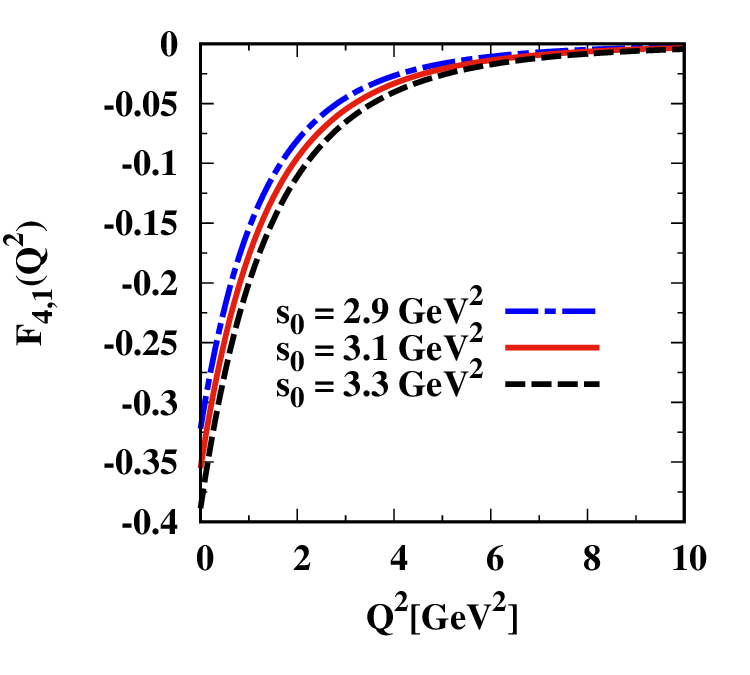}\\
	\vspace{-0.1cm}
	\includegraphics[width=0.35\textwidth]{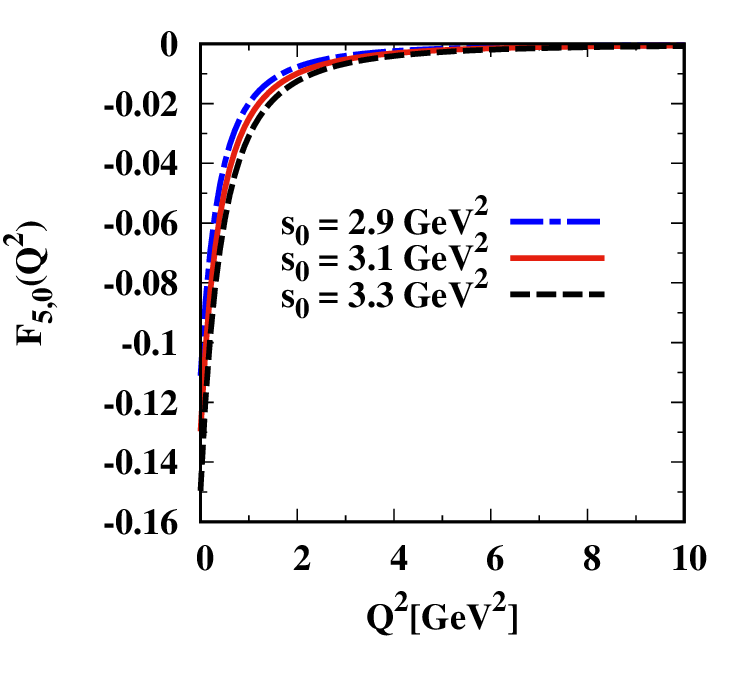}~~~~~~~~
	\caption{The dependence of the GFFs of $\Delta$ on $Q^2$ at $M^2$ = 2.5~GeV$^2$ for three values of $s_0$.}
	\label{fig:Qsqfigs}
\end{figure}

\begin{figure}[!htb]	
	\centering
	\includegraphics[width=0.35\textwidth]{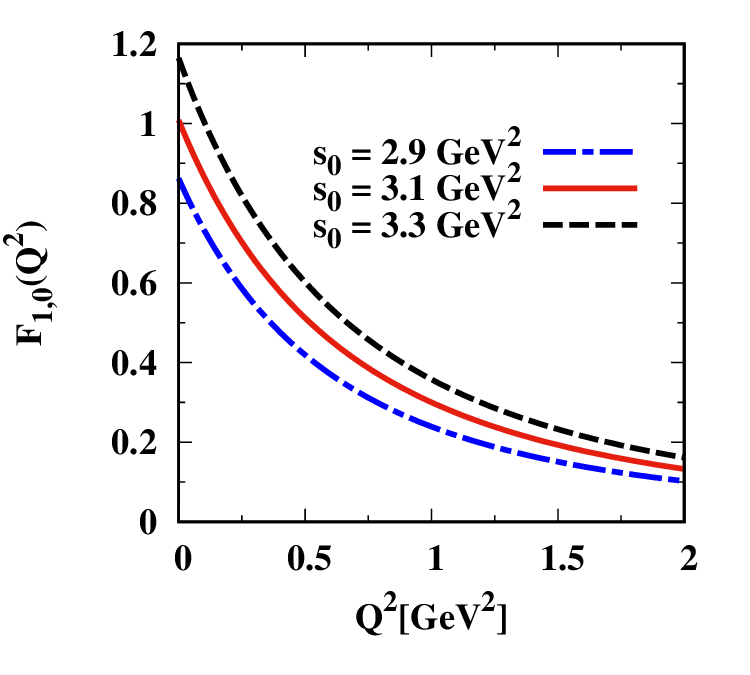}~~~~~~~~
	\includegraphics[width=0.35\textwidth]{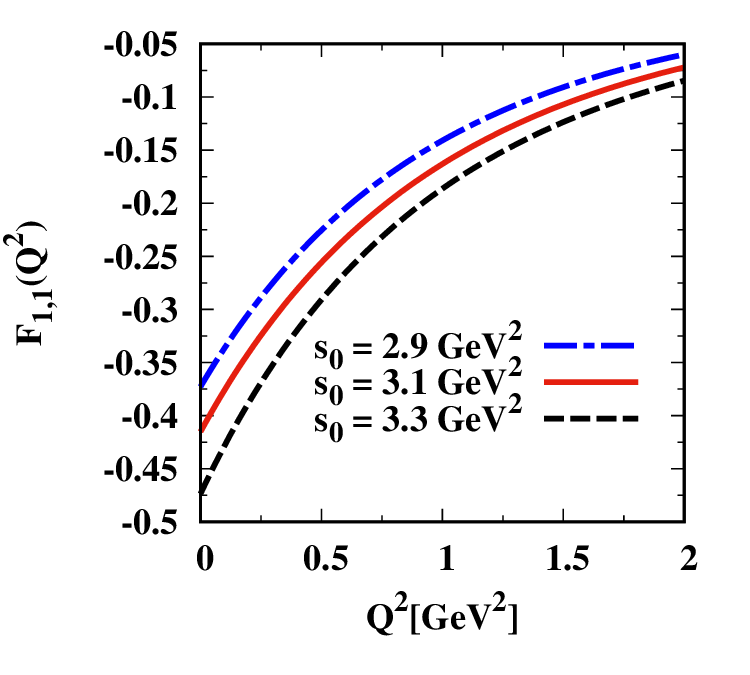}\\
	\vspace{-0.1cm}
	\includegraphics[width=0.35\textwidth]{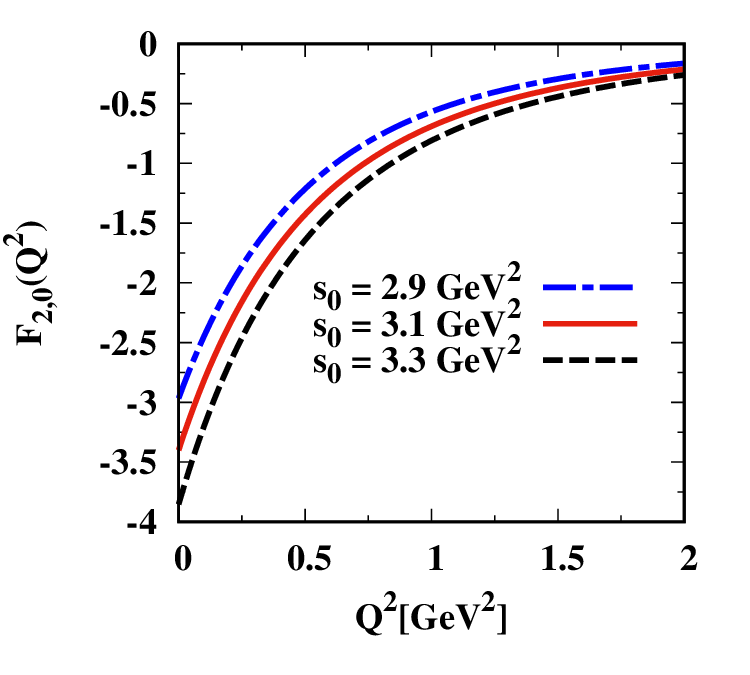}~~~~~~~~
	\includegraphics[width=0.35\textwidth]{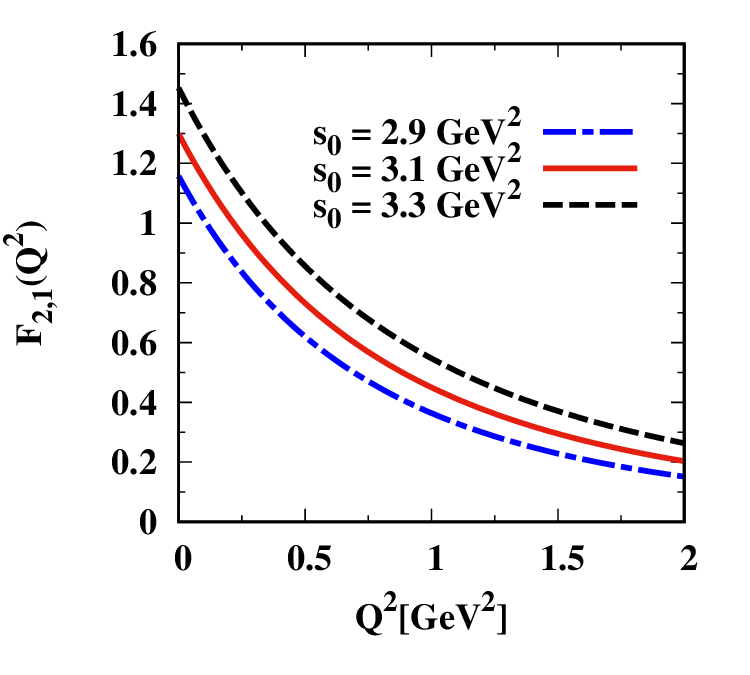}\\
	\vspace{-0.1cm}
	\includegraphics[width=0.35\textwidth]{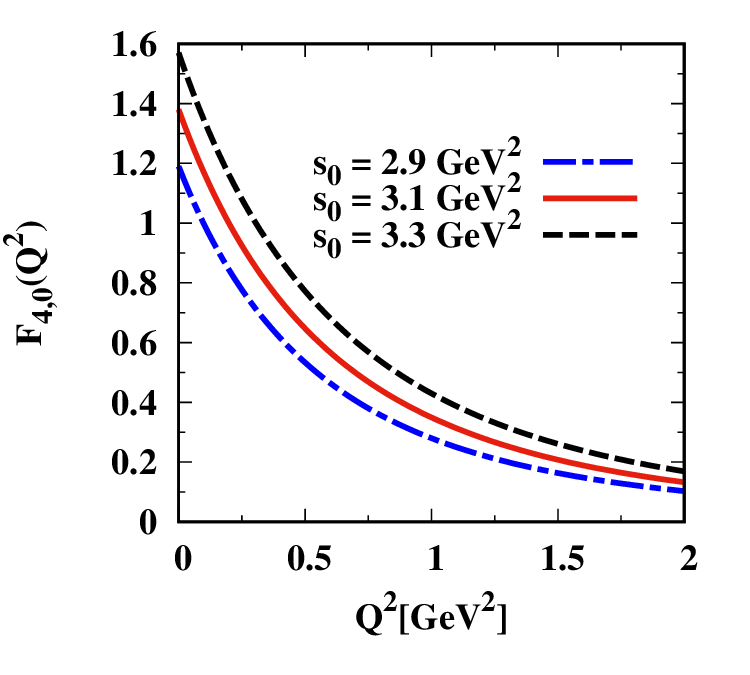}~~~~~~~~
	\includegraphics[width=0.35\textwidth]{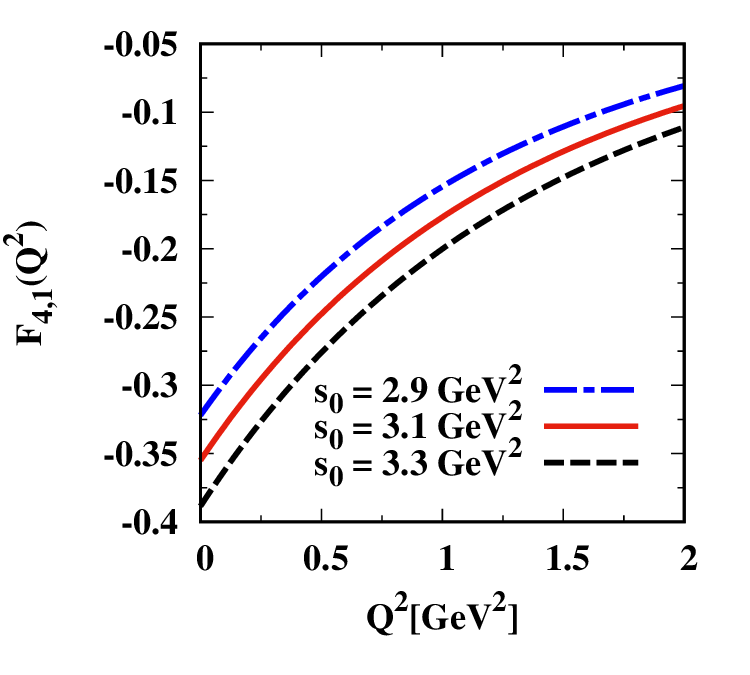}\\
	\vspace{-0.1cm}
	\includegraphics[width=0.35\textwidth]{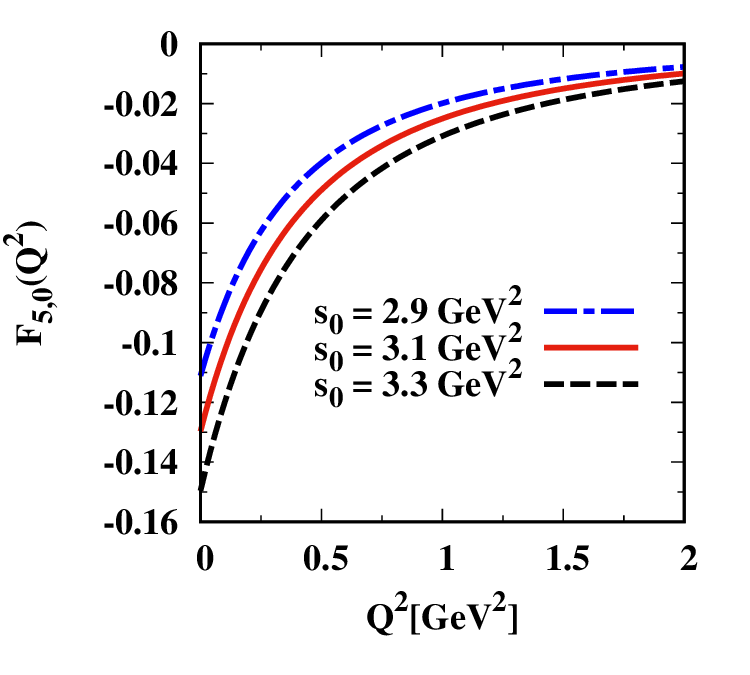}~~~~~~~~
	\caption{The $Q^2$ dependence of $\Delta$'s GFFs at $M^2$ = 2.5~GeV$^2$ for three values of $s_0$, where we restricted $0 \,{\text{GeV}^2} \leqslant {Q^2} \leqslant 2 \,{\text{GeV}^2}$ for more clarity.}
	\label{fig:smallQsqfigs}
\end{figure}

In Fig.~\ref{fig:Msqfigs}, we present the Borel mass parameter dependence of the GFFs at $Q^2 = 1.0$ GeV$^2$ and three values of the continuum threshold $s_0 = 2.9, 3.1$ and $3.3$ GeV$^2$. This figure shows that the GFFs are stable with respect to the change of Borel mass parameter in the working region.
In Fig.~\ref{fig:Qsqfigs}, we present the GFFs as a function of $Q^2$ for the fixed Borel mass $M^2 = 2.5$ GeV$^2$ and three values of the continuum threshold $s_0 = 2.9, 3.1$ and $3.3$ GeV$^2$. 
As expected, we observe that the $Q^2$ dependencies of the GFFs are smoothly changed and decrease with increasing  the $Q^2$. 
We use the following p-pole fit function to fit the GFFs from the sum rules predictions,
\begin{equation}\label{eq:FitFun}
{\cal F}(Q^2) = \frac{{\cal F}(0)}{\Big(1+ m_{p}\,Q^2\Big)^p},
\end{equation}
where the fit parameters ${\cal F} (0)$ and $p$ are dimensionless and $m_{p}$ has the inverse square energy dimension. The p-pole fit function of $\Delta$'s GFFs tends to zero at large $Q^2=10 ~{\text{GeV}^2}$, as Fig.~\ref{fig:Qsqfigs} illustrates. To enhance the visibility, Fig.~\ref{fig:smallQsqfigs} shows the $Q^2$ dependence of $\Delta$'s GFFs for $0 \,{\text{GeV}^2} \leqslant {Q^2} \leqslant 2 \,{\text{GeV}^2}$. The p-pole fit parameters of the GFFs in Fig.~\ref{fig:Qsqfigs} at mean values of the continuum threshold are summarized in Table.~\ref{table:fitparameters}. Changes in the working regions of auxiliary parameters, uncertainty in the input parameters as well as the systematic errors in QCD sum rules method cause errors in our presented results. Some mechanical properties  are revealed by the $\Delta$'s GFFs at zero momentum transfer, which are shown in the second column of this Table as ${\cal F} (0)$.

\begin{table}[!htb]
	\centering	
	\begin{tabular}{c*{12}{c}r}	
		\hline
  \\
		GFF & \qquad ${\cal F}(0)$  & \quad $m_{p}$~(GeV$^{-2}$) & 
		\quad $p$ \\
  \\
		\hline
		\hline
  \\
		$F_{1,0}(Q^2)$ & \qquad $ 1.01 \pm 0.15 $  &\quad $ 0.63 \pm 0.03$ &\quad $2.52 \pm 0.04$  \\
  \\
		$F_{1,1}(Q^2)$ & \qquad $-0.42 \pm 0.05$ &\quad$0.17 \pm 0.03$ &\quad$6.23 \pm 1.08$ \\
  \\
		$F_{2,0}(Q^2)$ & \qquad $-3.41 \pm 0.45$  &\quad$0.42 \pm 0.03$ &\quad $4.59 \pm 0.48$  \\
  \\
		$F_{2,1}(Q^2)$ & \qquad $1.30 \pm 0.15$  &\quad$0.38 \pm 0.01$ &\quad $3.27 \pm 0.40$  \\
  \\
		$F_{4,0}(Q^2)$ & \qquad $1.38 \pm 0.19$  &\quad$0.51 \pm 0.04$ &\quad $3.32 \pm 0.04$  \\
  \\
		$F_{4,1}(Q^2)$ & \qquad $-0.35 \pm 0.03$  &\quad$0.14 \pm 0.01$ &\quad $5.41 \pm 0.16$  \\
  \\
		$F_{5,0}(Q^2)$ & \qquad $-0.13 \pm 0.02$  &\quad$1.14 \pm 0.07$ &\quad $2.17 \pm 0.01$  \\
  \\
		\hline
	\end{tabular}
	\caption{ The numerical values of p-pole fit parameters ${\cal F}(0)$,  $m_{p}$ and $ p $ of the GFFs in Fig.~\ref{fig:Qsqfigs} at mean values of the continuum threshold. }
	\label{table:fitparameters}
\end{table}

We present and compare some mechanical properties extracted from our work and other studies in the rest of this section. Table.~\ref{table:CompareResultsZeroQ2} shows some of the GMFFs of $\Delta$ baryon at zero momentum transfer obtained from our calculations and compares them with the results of Ref.~\cite{Kim:2020lrs}. The normalization condition for $\Delta$ mass is $1$, which is consistent with $\varepsilon_0(0) = F_{1,0}(0) = 1.01 \pm 0.15$ from our calculations. We obtain $\mathcal{J}_1(0) = \frac{1}{3} F_{4,0}(0) =  0.46 \pm 0.06$ for the dipole angular momentum where $F_{4,0}(0) = 1.38 \pm 0.19$ corresponds to spin of $\Delta$ baryon which is $3/2$. This result is well consistent with the prediction of Skyrme model within the presented errors.
We obtain a p-pole behaviour for the octupole angular momentum form factor $\mathcal{J}_3(Q^2)$ from our calculations and Eq.~\eqref{eq:GMFFsJ3}. At zero momentum transfer, we have $\mathcal{J}_3(0) = -\frac{1}{6} [F_{4,0}(0) + F_{4,1}(0)] =  -0.17 \pm 0.03$ and at large momentum transfer $Q^2 = 10$, $\mathcal{J}_3(Q^2)$ approach to zero. In contrast, Ref.~\cite{Kim:2020lrs} assumes that $\mathcal{J}_3(Q^2)$ is zero for all values of $Q^2$ to suppress the corresponding density in the large $N_c$ expansion. Our obtained $\varepsilon_2(0) = -0.18 \pm 0.03$ differs from the corresponding value in Ref.~\cite{Kim:2020lrs}.

Except for $F_{1,1}(0)$ and $F_{4,1}(0)$, our results for $\Delta$'s GFFs at zero momentum transfer, ${\cal F}(0)$ in Table.~\ref{table:fitparameters}, are comparable with the corresponding results in Ref.~\cite{Kim:2020lrs}. We obtain $F_{1,1}(0) = -0.42 \pm 0.05$ and $F_{4,1}(0) = -0.35 \pm 0.03$ from our calculations, which contrast with $F_{1,1}(0) = -3.64$ and $F_{4,1}(0) = -1.5$ in Ref.~\cite{Kim:2020lrs}. By applying the Skyrme model with the constraints $\varepsilon_0(0) = F_{1,0}(0) = 1$ and $\mathcal{J}_1(0) = \frac{1}{3} F_{4,0}(0) = \frac{1}{2}$ and the assumption $\mathcal{J}_3(0) = -\frac{1}{6} [F_{4,0}(0) + F_{4,1}(0)] = 0$, Ref.~\cite{Kim:2020lrs} obtains $F_{4,1}(0) = -F_{4,0}(0)= -1.5$. The sum rules method allows us to define $\Delta$'s GFFs without imposing any additional conditions on GFFs and GMFFs, which is an advantage of this method. The ratio $F_{1,1}(0)/F_{4,1}(0)$ is obtained using our sum rules  and compared with some other models' predictions, as shown by,
\begin{equation}
\frac{F_{1,1}(0)}{F_{4,1}(0)} \simeq \begin{cases}
$2$ & \qquad \text{tree-level chiral perturbation theory (ChPT)~\cite{Alharazin:2022wjj}},\\
$2.43$ & \qquad \text{Skyrme model~\cite{Kim:2020lrs}},\\
\text{$1.2 \pm 0.25$} & \qquad \text{ current work}.
\end{cases}
\end{equation}
As is seen, the  different approaches agree on the sign  this ratio and the obtained magnitudes are roughly  close to each other. Note that it is not possible to extract the values for $F_{1,1}(0)$ and $F_{4,1}(0)$ using ChPT, because the needed coupling constants are not fixed (see Ref.~\cite{Alharazin:2022wjj}).

Our results for $D_2(0)$ and $D_3(0)$ agree with those of Ref.~\cite{Kim:2020lrs} while for $D_0(0)$ differs from the corresponding value in this reference considerably. 
The D-terms and the mass radius of our calculations for the $\Delta$ baryon are shown in Table.~\ref{table:CompareResults} along with the predictions of other models. We get $\langle r^2_E\rangle = 0.67 \pm 0.04$ fm$^2$ for the mass radius from Eq.~\eqref{eq:radius}, which agrees, within the uncertainties of our result,  with the $0.64$ fm$^2$ reported in Refs.~\cite{Perevalova:2016dln, Kim:2020lrs}. 
While our result for $\mathcal{D}_0^{\Delta}$ is consistent with that of  Ref. \cite{Panteleeva:2020ejw}, it is quite different from the prediction of Ref. \cite{Kim:2020lrs} result. When the $\mathcal{D}_2^{\Delta}$ and $\mathcal{D}_3^{\Delta}$ results are examined, it is seen that our results are compatible with the results of Refs.~\cite{Panteleeva:2020ejw, Kim:2020lrs} within the errors.
The vanishing of $\mathcal{D}_2^{\Delta}$ in the QCD sum rule approach is significant, as it confirms the remarkable prediction of viewing baryons as the chiral solitons \cite{Panteleeva:2020ejw}.

\begin{table}[!htb]
\centering	
\begin{tabular}{c*{12}{c}r}	
\hline\hline
\\
\textbf{Model} &\quad $\varepsilon_0(0)$ &\quad 
$
\varepsilon_2(0) 
$
&\quad
$\mathcal{J}_1(0)$ &\quad 
$
\mathcal{J}_3(0)
$
&\quad
$D_0(0)$ &\quad 
$D_2(0)$ &\quad $D_3(0)$  \\
\\
\hline
\hline
\\
This Work  &\quad $ 1.01 \pm 0.15 $ &\quad $-0.18 \pm 0.03$ &\quad
$0.46 \pm 0.06$ &\quad
$-0.17 \pm 0.03$ &\quad
$-2.71 \pm 0.34$ &\quad $-0.17 \pm 0.03$ &\quad $0.26 \pm 0.04$ 
 \\
 \\
\hline
\\
\cite{Kim:2020lrs} &\quad $1$ &\quad $0.34$ &\quad $0.5$ &\quad $0$ &\quad $-3.53$
&\quad $-0.20$ &\quad $0.24$   \\
\\
\hline\hline
\end{tabular}
\caption{ A comparison of mechanical properties obtained in the present study at zero momentum transfer with those from Skyrme model~\cite{Kim:2020lrs}. }
\label{table:CompareResultsZeroQ2}
\end{table}

By means of the $\Delta$ baryon's D-terms, one can obtain the D-term of the nucleon using the large $ N_c $ picture of baryons as chiral solitons  as follows~\cite{Panteleeva:2020ejw}:
\begin{equation}\label{eq:DtermDelta}
 \mathcal{D}_0^{N} = \mathcal{D}_0^{\Delta} + 2 \mathcal{D}_3^{\Delta}.
\end{equation}
The above relation yields $\mathcal{D}_0^{N} = -3.57 \pm 0.46$ using the D-terms of the $\Delta$ baryon in our calculations, which is in good agreement with the $\mathcal{D}_0^{N}$ values of Refs.~\cite{Panteleeva:2020ejw, Kim:2020lrs}. 
From these results, we see that the D-term $\mathcal{D}_0^{N}$ of nucleon has a higher absolute value than the generalized D-terms $\mathcal{D}_0^{\Delta}$ and $\mathcal{D}_3^{\Delta}$ of the $\Delta$ baryon, which are all negative as expected:  It is thought that if a system satisfies the local stability conditions, the D-terms should be negative, if not the system would collapse.

\begin{table}[!htb]
\small
\centering	
\begin{tabular}{c*{12}{c}r}	
\hline\hline
\\
\textbf{Model} &\quad $\mathcal{D}_0^{\Delta}$ &\quad  $\mathcal{D}_2^{\Delta}$ &\quad $\mathcal{D}_3^{\Delta}$ &\quad $\mathcal{D}_0^{N}$ &\quad $\langle r^2_E\rangle ~ (\text{fm}^2)$ 
\\
\\
\hline
\hline
\\
This Work &\quad $-2.71 \pm 0.34$ &\quad $0.000 \pm 0.002$ &\quad $-0.43 \pm 0.06$ & \quad $-3.57 \pm 0.46$ &\quad $0.67 \pm 0.04$  \\
\\
\hline
\\
\cite{Perevalova:2016dln, Panteleeva:2020ejw} &\quad $-2.65$ &\quad $0$ &\quad $-0.38$ & \quad $-3.40$ &\quad $0.64$ \\
\\
\hline
\\
\cite{Kim:2020lrs} &\quad $-3.53$ &\quad $0$ &\quad $-0.50$ & \quad $-3.63$ &\quad $0.64$ 
\\
\\
\hline\hline
\end{tabular}
\caption{ A comparison of the D-terms and the mass radius obtained in the present study with those from Skyrme model \cite{Perevalova:2016dln, Panteleeva:2020ejw, Kim:2020lrs}.
}
\label{table:CompareResults}
\end{table}

\section{Summary and conclusion}\label{sec:conclusion}
Due to the different interaction types, a hadron can have different kinds of form factors representing the corresponding interaction. Determination of different form factors of hadrons  allow us  to obtain useful information about the various related physical quantities that can help us discover the nature and internal structures of hadrons as well as the  nonperturbative nature of  QCD as the theory of  strong interaction. The gravitational form factors that emerge as a result of the graviton-like interaction of the hadrons with the energy-momentum tensor current are of great importance as they provide important information about the inner structures,  quark-gluon organizations of hadrons,  distributions of the strong forces, energy and  pressure inside them as well as their geometric shape and radius. These cause an increasing  interest to investigation of hadronic GFFs.

In this study, we investigated the $\Delta \to \Delta$ transition in the presence of the energy-momentum tensor current. We considered both the quark and gluonic parts of the EMT current.  Such interaction is parameterized in terms of  ten GFFs: Seven conserved and three non-conserved form factors. The non-conserved form factors vanish because of the conservation of the total EMT current. We derived the sum rules and numerically determined the seven conserved GFFs  of the $\Delta$ baryon in the range $0 \leq Q^2 \leq 10$ GeV$^2$ using the three-point QCD sum rules approach. The QCD sum rule method is a relativistic method and considers different features and quantum numbers of the hadrons like their spin, being one of the leading existing nonperturbaative approaches. We found that the $Q^2$ behavior  of $\Delta$'s GFFs  are well explained via a p-pole fit function. We presented the values of the GFFs at zero momentum transfer as well. 

Having determined the GFFs of the $\Delta$ baryon,  we used them to calculate the composite gravitational form factors  of the system  like the energy and angular momentum multipole form factors, $ \mathcal{D} $ terms representing the  mechanical properties like the internal pressure and shear forces as well as the mass radius of $ \Delta$ resonance and compared them with other existing theoretical predictions. Our results obtained using QCD sum rules agree with the remarkable prediction of the soliton picture of baryons, which resulted  from vanishing of $\mathcal{D}_2^{\Delta}$ term.

Our results on $\varepsilon_0 (Q^2)$, $\varepsilon_2 (Q^2)$, $\mathcal{J}_1 (Q^2)$ and $\mathcal{J}_3 (Q^2)$, which  are respectively energy-monopole, energy-quadrupole, angular momentum-dipole and angular momentum-octupole form factors as well as  $D_{0, 2, 3} (Q^2)$  composite form factors related to the internal pressures and shear forces  and the generalized D-terms $\mathcal{D}_{0,2,3}$  satisfy the required conditions and describe well different features of the $\Delta$ baryon. Our results may be compared with future probable lattice QCD and other theoretical predictions. We hope that such investigations will be possible in the future experiments as well. If the direct measurements of the quantities considered in the present study are difficult because of the short lifetime of the $\Delta$ baryon, we hope that we can extract GPDs of this system using experimental data on different related  physical quantities like electromagnetic form factors and multipole moments. As we previously mentioned, one can determine the GFFs using the extracted GPDs from the experimental data. Comparison of the obtained GFFs by this way with the results of the present study will be of great importance as was done for the nucleon in Ref.  \cite{Hashamipour:2022noy}.

\appendix
\section{QCD side solutions of three-point correlation function} 
\label{Calcu}

\setcounter{equation}{0}
\renewcommand{\theequation}{\Alph{section}.\arabic{equation}}

In this appendix, we collect some parts of QCD side solutions of the three-point correlation function. In Eq.~\eqref{eq:contractions}, after using Wick's theorem and calculating all possible contractions, $\Gamma^q$ and $\Gamma^g$ are obtained as below,
\begin{align}\label{eq:qcontractions}
	\begin{aligned}
		\Gamma_{\alpha\mu\nu\beta}^q = i 
		& \Big\{ 
		4 S^{cc'}_{u}(y-x) Tr \Big[ \gamma_{\beta} S'^{bb'}_{d}(y-x) \gamma_{\alpha} S^{am}_{u}(y-z) \gamma_{\nu}
		\overleftrightarrow{D}_\mu (z) S^{ma'}_{u}(z-x) \Big] \\
		&-4 S^{ca'}_{u}(y-x) \gamma_{\beta}
		S'^{bb'}_{d}(y-x) \gamma_{\alpha}
		S^{am}_{u}(y-z) \gamma_{\nu}
		\overleftrightarrow{D}_\mu (z) S^{mc'}_{u}(z-x)\\
		&-4 S^{cm}_{u}(y-z) \gamma_{\nu}
		\overleftrightarrow{D}_\mu (z) S^{ma'}_{u}(z-x)
		\gamma_{\beta} S'^{bb'}_{d}(y-x) \gamma_{\alpha}S^{ac'}_{u}(y-x) \\ 
		&+4 S^{cm}_{u}(y-z) \gamma_{\nu}
		\overleftrightarrow{D}_\mu (z) S^{mc'}_{u}(z-x)
		Tr \Big[ \gamma_{\beta} S'^{bb'}_{d}(y-x) \gamma_{\alpha}S^{aa'}_{u}(y-x) \Big]\\
		&+4 S^{cc'}_{u}(y-x) Tr \Big[ \gamma_{\beta}
		S'^{aa'}_{u}(y-x) \gamma_{\alpha} S^{bm}_{d}(y-z) \gamma_{\nu}
		\overleftrightarrow{D}_\mu (z) S^{mb'}_{d}(z-x) \Big]\\
		&-4 S^{ca'}_{u}(y-x) \gamma_{\beta}
		S'^{mb'}_{d}(z-x) \overleftrightarrow{D}_\mu (z) 
		\gamma_{\nu} S'^{bm}_{d}(y-z) \gamma_{\alpha} S^{ac'}_{u}(y-x) \\
		&+ 2 S^{ca'}_{u}(y-x) \gamma_{\beta}
		S'^{mb'}_{u}(z-x) \overleftrightarrow{D}_\mu (z) 
		\gamma_{\nu} S'^{am}_{u}(y-z) \gamma_{\alpha} S^{bc'}_{d}(y-x) \\
		&- 2 S^{cb'}_{u}(y-x) \gamma_{\beta}
		S'^{ma'}_{u}(z-x) \overleftrightarrow{D}_\mu (z) 
		\gamma_{\nu} S'^{am}_{u}(y-z) \gamma_{\alpha} S^{bc'}_{d}(y-x) \\
		&+2 S^{cm}_{u}(y-z) \gamma_{\nu}
		\overleftrightarrow{D}_\mu (z) S^{ma'}_{u}(z-x)
		\gamma_{\beta} S'^{ab'}_{u}(y-x) \gamma_{\alpha}S^{bc'}_{d}(y-x) \\ 
		&-2 S^{cm}_{u}(y-z) \gamma_{\nu}
		\overleftrightarrow{D}_\mu (z) S^{mb'}_{u}(z-x)
		\gamma_{\beta} S'^{aa'}_{u}(y-x) \gamma_{\alpha}S^{bc'}_{d}(y-x) \\
		&+2 S^{ca'}_{u}(y-x) \gamma_{\beta}
		S'^{ab'}_{u}(y-x) \gamma_{\alpha}
		S^{bm}_{d}(y-z) \gamma_{\nu}
		\overleftrightarrow{D}_\mu (z) S^{mc'}_{d}(z-x)\\
		&-2 S^{cb'}_{u}(y-x) \gamma_{\beta}
		S'^{aa'}_{u}(y-x) \gamma_{\alpha}
		S^{bm}_{d}(y-z) \gamma_{\nu}
		\overleftrightarrow{D}_\mu (z) S^{mc'}_{d}(z-x)\\
		&+2 S^{cb'}_{d}(y-x) \gamma_{\beta}
		S'^{ba'}_{u}(y-x) \gamma_{\alpha}
		S^{am}_{u}(y-z) \gamma_{\nu}
		\overleftrightarrow{D}_\mu (z) S^{mc'}_{u}(z-x)\\
		&-2 S^{cb'}_{d}(y-x) \gamma_{\beta}
		S'^{ma'}_{u}(z-x) \overleftrightarrow{D}_\mu (z) 
		\gamma_{\nu} S'^{am}_{u}(y-z) \gamma_{\alpha} S^{bc'}_{u}(y-x) \\
		&+2 S^{cb'}_{d}(y-x) \gamma_{\beta}
		S'^{ma'}_{u}(z-x) \overleftrightarrow{D}_\mu (z) 
		\gamma_{\nu} S'^{bm}_{u}(y-z) \gamma_{\alpha} S^{ac'}_{u}(y-x) \\
		&-2 S^{cb'}_{d}(y-x) \gamma_{\beta}
		S'^{aa'}_{u}(y-x) \gamma_{\alpha}
		S^{bm}_{u}(y-z) \gamma_{\nu}
		\overleftrightarrow{D}_\mu (z) S^{mc'}_{u}(z-x)\\
		&+2 S^{cm}_{d}(y-z) \gamma_{\nu}
		\overleftrightarrow{D}_\mu (z) S^{mb'}_{d}(z-x)
		\gamma_{\beta} S'^{ba'}_{u}(y-x) \gamma_{\alpha}S^{ac'}_{u}(y-x) \\
		&-2 S^{cm}_{d}(y-z) \gamma_{\nu}
		\overleftrightarrow{D}_\mu (z) S^{mb'}_{d}(z-x)
		\gamma_{\beta} S'^{aa'}_{u}(y-x) \gamma_{\alpha}S^{bc'}_{u}(y-x) \\
		&+ S^{cc'}_{d}(y-x) Tr \Big[ \gamma_{\beta} S'^{bb'}_{u}(y-x) \gamma_{\alpha} S^{am}_{u}(y-z) \gamma_{\nu}
		\overleftrightarrow{D}_\mu (z) S^{ma'}_{u}(z-x) \Big] \\
		&- S^{cc'}_{d}(y-x) Tr \Big[ \gamma_{\beta} S'^{ba'}_{u}(y-x) \gamma_{\alpha} S^{am}_{u}(y-z) \gamma_{\nu}
		\overleftrightarrow{D}_\mu (z) S^{mb'}_{u}(z-x) \Big] \\
		&- S^{cc'}_{d}(y-x) Tr \Big[ \gamma_{\beta} S'^{ab'}_{u}(y-x) \gamma_{\alpha} S^{bm}_{u}(y-z) \gamma_{\nu}
		\overleftrightarrow{D}_\mu (z) S^{ma'}_{u}(z-x) \Big] \\
		&+ S^{cc'}_{d}(y-x) Tr \Big[ \gamma_{\beta} S'^{aa'}_{u}(y-x) \gamma_{\alpha} S^{bm}_{u}(y-z) \gamma_{\nu}
		\overleftrightarrow{D}_\mu (z) S^{mb'}_{u}(z-x) \Big] \\
		&+ S^{cm}_{d}(y-z) \gamma_{\nu}
		\overleftrightarrow{D}_\mu (z) S^{mc'}_{d}(z-x)
		Tr \Big[ \gamma_{\beta} S'^{aa'}_{u}(y-x) \gamma_{\alpha}S^{bb'}_{u}(y-x) \Big]\\
		&- S^{cm}_{d}(y-z) \gamma_{\nu}
		\overleftrightarrow{D}_\mu (z) S^{mc'}_{d}(z-x)
		Tr \Big[ \gamma_{\beta} S'^{ab'}_{u}(y-x) \gamma_{\alpha}S^{ba'}_{u}(y-x) \Big]
		+ \mu \leftrightarrow \nu \Big\},\\ 
	\end{aligned}
\end{align}

\begin{align}\label{eq:gcontractions}
	\begin{aligned}
		\Gamma^g_{\alpha\mu\nu\beta} &= \langle G^2 \rangle g_{\mu\nu} 
		\Big\{ 
		4 S^{cc'}_{u}(y-x) Tr \Big[ \gamma_{\beta} S'^{bb'}_{d}(y-x)
		\gamma_{\alpha}S^{aa'}_{u}(y-x) \Big]
		-4 S^{ca'}_{u}(y-x) \gamma_{\beta} S'^{bb'}_{d}(y-x)
		\gamma_{\alpha}S^{ac'}_{u}(y-x)  \\
		&+2 S^{ca'}_{u}(y-x) \gamma_{\beta} S'^{ab'}_{u}(y-x)
		\gamma_{\alpha}S^{bc'}_{d}(y-x) 
		-2 S^{cb'}_{u}(y-x) \gamma_{\beta} S'^{aa'}_{u}(y-x)
		\gamma_{\alpha}S^{bc'}_{d}(y-x)  \\
		&+2 S^{cb'}_{d}(y-x) \gamma_{\beta} S'^{ba'}_{u}(y-x)
		\gamma_{\alpha}S^{ac'}_{u}(y-x) 
		-2 S^{cb'}_{d}(y-x) \gamma_{\beta} S'^{aa'}_{u}(y-x)
		\gamma_{\alpha}S^{bc'}_{u}(y-x)  \\
		&+ S^{cc'}_{d}(y-x) Tr \Big[ \gamma_{\beta} S'^{aa'}_{u}(y-x)
		\gamma_{\alpha}S^{bb'}_{u}(y-x) \Big] 
		- S^{cc'}_{d}(y-x) Tr \Big[ \gamma_{\beta} S'^{ab'}_{u}(y-x)
		\gamma_{\alpha}S^{ba'}_{u}(y-x) \Big] 
		\Big\},
	\end{aligned}
\end{align}

where $S' = C S^T C$ and $S^{ij}_q(x)$ is the light quark propagator, defined by,
\begin{align}\label{eq:lightquarkProp}
	S^{ij}_q(x) &= i \delta_{ij} \dfrac{\xslash}{2\pi^2 x^4} - 
	\delta_{ij} \dfrac{m_q}{4 \pi^2 x^2} - \delta_{ij} \dfrac{\langle\bar{q}q\rangle}{12} 
	+ i \delta_{ij} \dfrac{\xslash m_q \langle\bar{q}q\rangle}{48}
	- \delta_{ij} \dfrac{x^2}{192} m_0^2 \langle\bar{q}q\rangle 
	+ i \delta_{ij} \dfrac{x^2 \xslash m_q}{1152} m_0^2 \langle\bar{q}q\rangle 
 \nonumber\\
 &
	- i \dfrac{g_s G_{ij}^{\lambda\delta}}{32 \pi^2 x^2} 
	[\xslash \sigma_{\lambda\delta} + \sigma_{\lambda\delta} \xslash]
	+...
\end{align}
with $m_0^2 = \langle\bar{q}g_s G^{\mu\nu}\sigma_{\mu\nu}q\rangle / \langle\bar{q}q\rangle$ and we assume $m_q=0$. 

The perturbative and non-perturbative contributions of the correlation function in Eq.~\eqref{eq:totalGamma} are given by,
\begin{widetext} 
	\begin{align}\label{eq:perturbative}
		\begin{aligned}
			\Gamma_{\alpha\mu\nu\beta}^{(P)} &=
			\dfrac{3 i^7}{(2 \pi^2)^4} \dfrac{1}{(y-x)^{8}}
			\Big\{ 2 (\yslash - \xslash) 
			Tr \Big[ \gamma_{\beta} (\yslash - \xslash) 
			\gamma_{\alpha} A^P_{\mu\nu} (x,y) \Big]
			+ 2 (\yslash - \xslash) \gamma_{\beta} (\yslash - \xslash) 
			\gamma_{\alpha} A^P_{\mu\nu} (x,y) \\
			&+ 2 A^P_{\mu\nu} (x,y) \gamma_{\beta} (\yslash - \xslash)
			\gamma_{\alpha} (\yslash - \xslash)
			+  A^P_{\mu\nu} (x,y)
			Tr \Big[ \gamma_{\beta} (\yslash - \xslash) 
			\gamma_{\alpha} (\yslash - \xslash) \Big] 
			+ 2 (\yslash - \xslash) \gamma_{\beta}
			B^P_{\mu\nu} (x,y) \gamma_{\alpha} (\yslash - \xslash) \Big\},
		\end{aligned}
	\end{align}
\end{widetext}

\begin{widetext} 
\begin{align}\label{eq:3D}
\begin{aligned}
\hspace{-0.226 cm}\Gamma_{\alpha\mu\nu\beta}^{(3D)} &=
			\dfrac{i^6}{4 (2 \pi^2)^3} \dfrac{\langle \bar{q} q \rangle}{(y-x)^{4}}
			\Big\{ 2 (\yslash - \xslash) 
			Tr \Big[ \gamma_{\beta} \gamma_{\alpha} A^P_{\mu\nu} (x,y) \Big]
			- 2  Tr \Big[ \gamma_{\beta} (\yslash - \xslash) 
			\gamma_{\alpha} A^P_{\mu\nu} (x,y) \Big]
			+ 2 (\yslash - \xslash) \gamma_{\beta} \gamma_{\alpha} A^P_{\mu\nu} (x,y) \\
			&- 2 \gamma_{\beta} (\yslash - \xslash) \gamma_{\alpha} A^P_{\mu\nu} (x,y)
			+ 2 A^P_{\mu\nu} (x,y) \gamma_{\beta} \gamma_{\alpha} (\yslash - \xslash)
			- 2 A^P_{\mu\nu} (x,y) \gamma_{\beta} (\yslash - \xslash) \gamma_{\alpha} 
			+ A^P_{\mu\nu} (x,y) Tr \Big[ \gamma_{\beta}  
			\gamma_{\alpha} (\yslash - \xslash) \Big] \\
			&- A^P_{\mu\nu} (x,y) Tr \Big[ \gamma_{\beta} (\yslash - \xslash) 
			\gamma_{\alpha} \Big] 
			- 2 (\yslash - \xslash) \gamma_{\beta}
			B^P_{\mu\nu} (x,y) \gamma_{\alpha}  
			- 2  \gamma_{\beta} B^P_{\mu\nu} (x,y) \gamma_{\alpha} (\yslash - \xslash) \\
			&- \dfrac{1}{(y-x)^{4}}
			\Big(
			2 (\yslash - \xslash) 
			Tr \Big[ \gamma_{\beta} (\yslash - \xslash) 
			\gamma_{\alpha} A^3_{\mu\nu} (x,y) \Big]
			+ 2 (\yslash - \xslash) \gamma_{\beta} (\yslash - \xslash) 
			\gamma_{\alpha} A^3_{\mu\nu} (x,y) \\
			&+ 2 A^3_{\mu\nu} (x,y) \gamma_{\beta} (\yslash - \xslash)
			\gamma_{\alpha} (\yslash - \xslash)
			+  A^3_{\mu\nu} (x,y)
			Tr \Big[ \gamma_{\beta} (\yslash - \xslash) 
			\gamma_{\alpha} (\yslash - \xslash) \Big] 
			+ 2 (\yslash - \xslash) \gamma_{\beta}
			B^3_{\mu\nu} (x,y) \gamma_{\alpha} (\yslash - \xslash)
			\Big) \Big\},
		\end{aligned}
	\end{align}
\end{widetext}

\begin{widetext} 
	\begin{align}\label{eq:4Dquark}
		\begin{aligned}
			\Gamma_{\alpha\mu\nu\beta}^{(4D,q)} &=
			\dfrac{i^7}{24 (8 \pi^2)^4} \dfrac{g_s^2 \langle G^2 \rangle}{(y-x)^{4}}
			\Big\{ 2 \Big( (\yslash - \xslash)\sigma^{\lambda\delta}
			+ \sigma^{\lambda\delta} (\yslash - \xslash) \Big)
			Tr \Big[ \gamma_{\beta} 
			\Big( (\yslash - \xslash)\sigma_{\lambda\delta}
			+ \sigma_{\lambda\delta} (\yslash - \xslash) \Big)
			\gamma_{\alpha} A^P_{\mu\nu} (x,y) \Big]
			\\ 
			&+ 2 \Big( (\yslash - \xslash)\sigma^{\lambda\delta}
			+ \sigma^{\lambda\delta} (\yslash - \xslash) \Big)
			\gamma_{\beta} 
			\Big( (\yslash - \xslash)\sigma_{\lambda\delta}
			+ \sigma_{\lambda\delta} (\yslash - \xslash) \Big)
			\gamma_{\alpha} A^P_{\mu\nu} (x,y)
			\\
			&+ 2 A^P_{\mu\nu} (x,y) \gamma_{\beta}
			\Big( (\yslash - \xslash)\sigma^{\lambda\delta}
			+ \sigma^{\lambda\delta} (\yslash - \xslash) \Big)
			\gamma_{\alpha}  
			\Big( (\yslash - \xslash)\sigma_{\lambda\delta}
			+ \sigma_{\lambda\delta} (\yslash - \xslash) \Big)
			\\
			&+ A^P_{\mu\nu} (x,y) Tr \Big[ \gamma_{\beta} 
			\Big( (\yslash - \xslash)\sigma^{\lambda\delta}
			+ \sigma^{\lambda\delta} (\yslash - \xslash) \Big)
			\gamma_{\alpha}  
			\Big( (\yslash - \xslash)\sigma_{\lambda\delta}
			+ \sigma_{\lambda\delta} (\yslash - \xslash) \Big) \Big]
			\\
			&- 2 \Big( (\yslash - \xslash)\sigma^{\lambda\delta}
			+ \sigma^{\lambda\delta} (\yslash - \xslash) \Big)
			\gamma_{\beta} B^P_{\mu\nu} (x,y) \gamma_{\alpha} 
			\Big( (\yslash - \xslash)\sigma_{\lambda\delta}
			+ \sigma_{\lambda\delta} (\yslash - \xslash) \Big)  \\
			&+ \dfrac{1}{2 (y-x)^{4}}
			\Big(
			2 (\yslash - \xslash) 
			Tr \Big[ \gamma_{\beta} (\yslash - \xslash)
			\gamma_{\alpha}  A^{G^2}_{\mu\nu} (x,y)
			\Big]
			+ 2 (\yslash - \xslash)  \gamma_{\beta} (\yslash - \xslash)
			\gamma_{\alpha}  A^{G^2}_{\mu\nu} (x,y) \\
			&+ 2 A^{G^2}_{\mu\nu} (x,y)  \gamma_{\beta} (\yslash - \xslash)
			\gamma_{\alpha} (\yslash - \xslash)
			+ A^{G^2}_{\mu\nu} (x,y) Tr \Big[ \gamma_{\beta} (\yslash - \xslash)
			\gamma_{\alpha} (\yslash - \xslash) \Big]
			+ 2 (\yslash - \xslash) \gamma_{\beta}
			B^{G^2}_{\mu\nu} (x,y)  
			\gamma_{\alpha} (\yslash - \xslash) \Big)\\
			&+ \dfrac{1}{(y-x)^{2}}
			\Big( 2 (\yslash - \xslash)
			Tr \Big[ \gamma_{\beta} \Big( (\yslash - \xslash)\sigma^{\lambda\delta}
			+ \sigma^{\lambda\delta} (\yslash - \xslash) \Big)
			\gamma_{\alpha} A^{PG}_{\mu\nu} (x,y) \Big]\\
			&- 2 \Big( (\yslash - \xslash)\sigma^{\lambda\delta}
			+ \sigma^{\lambda\delta} (\yslash - \xslash) \Big)
			Tr \Big[ \gamma_{\beta} (\yslash - \xslash) 
			\gamma_{\alpha} A^{PG}_{\mu\nu} (x,y) \Big]\\
			&+ 2 (\yslash - \xslash)
			\gamma_{\beta} \Big( (\yslash - \xslash)\sigma^{\lambda\delta}
			+ \sigma^{\lambda\delta} (\yslash - \xslash) \Big)
			\gamma_{\alpha} A^{PG}_{\mu\nu} (x,y)
			- 2 \Big( (\yslash - \xslash)\sigma^{\lambda\delta}
			+ \sigma^{\lambda\delta} (\yslash - \xslash) \Big)
			\gamma_{\beta} (\yslash - \xslash) 
			\gamma_{\alpha} A^{PG}_{\mu\nu} (x,y) \\
			&+ 2 A^{PG}_{\mu\nu} (x,y) \gamma_{\beta}
			\Big( (\yslash - \xslash)\sigma^{\lambda\delta}
			+ \sigma^{\lambda\delta} (\yslash - \xslash) \Big)
			\gamma_{\alpha} (\yslash - \xslash)
			- 2 A^{PG}_{\mu\nu} (x,y) \gamma_{\beta} (\yslash - \xslash) \gamma_{\alpha}
			\Big( (\yslash - \xslash)\sigma^{\lambda\delta}
			+ \sigma^{\lambda\delta} (\yslash - \xslash) \Big) \\
			&+ A^{PG}_{\mu\nu} (x,y) \Big(
			Tr \Big[ \gamma_{\beta} \Big( (\yslash - \xslash)\sigma^{\lambda\delta}
			+ \sigma^{\lambda\delta} (\yslash - \xslash) \Big)
			\gamma_{\alpha} (\yslash - \xslash) \Big]
			- Tr \Big[ \gamma_{\beta} (\yslash - \xslash) \gamma_{\alpha}
			\Big( (\yslash - \xslash)\sigma^{\lambda\delta}
			+ \sigma^{\lambda\delta} (\yslash - \xslash) \Big) \Big]  \Big)\\
			&+ 2 (\yslash - \xslash) \gamma_{\beta} B^{PG}_{\mu\nu} (x,y) \gamma_{\alpha}
			\Big( (\yslash - \xslash)\sigma^{\lambda\delta}
			+ \sigma^{\lambda\delta} (\yslash - \xslash) \Big) 
			+ 2 \Big( (\yslash - \xslash)\sigma^{\lambda\delta}
			+ \sigma^{\lambda\delta} (\yslash - \xslash) \Big) \gamma_{\beta}
			B^{PG}_{\mu\nu} (x,y) \gamma_{\alpha} (\yslash - \xslash) \Big\}, 
		\end{aligned}
	\end{align}
\end{widetext}

\begin{widetext} 
\begin{equation}\label{eq:4Dgluon}	
\hspace{-3.2 cm}
\Gamma_{\alpha\mu\nu\beta}^{(4D,g)} =
	\dfrac{6 i^5}{(2 \pi^2)^3} \dfrac{\langle G^2 \rangle g_{\mu\nu}}{(y-x)^{12}}
	\Big\{ (\yslash - \xslash) 
	Tr \Big[ \gamma_{\beta} (\yslash - \xslash)
	\gamma_{\alpha} (\yslash - \xslash) \Big]
	+ 2 (\yslash - \xslash)  \gamma_{\beta} (\yslash - \xslash)
	\gamma_{\alpha} (\yslash - \xslash) \Big\},
	\end{equation}
\end{widetext}

\begin{widetext} 
	\begin{align}\label{eq:5D}
		\begin{aligned}
			\Gamma_{\alpha\mu\nu\beta}^{(5D)} &=
			\dfrac{i^6}{ (8 \pi^2)^3} \dfrac{m_0^2\langle \bar{q} q \rangle}{(y-x)^{2}}
			\Big\{ 2 (\yslash - \xslash) 
			Tr \Big[ \gamma_{\beta} \gamma_{\alpha} A^P_{\mu\nu} (x,y) \Big]
			- 2  Tr \Big[ \gamma_{\beta} (\yslash - \xslash) 
			\gamma_{\alpha} A^P_{\mu\nu} (x,y) \Big]
			+ 2 (\yslash - \xslash) \gamma_{\beta} \gamma_{\alpha} A^P_{\mu\nu} (x,y) \\
			&- 2 \gamma_{\beta} (\yslash - \xslash) \gamma_{\alpha} A^P_{\mu\nu} (x,y)
			+ 2 A^P_{\mu\nu} (x,y) \gamma_{\beta} \gamma_{\alpha} (\yslash - \xslash)
			- 2 A^P_{\mu\nu} (x,y) \gamma_{\beta} (\yslash - \xslash) \gamma_{\alpha} 
			+ A^P_{\mu\nu} (x,y) Tr \Big[ \gamma_{\beta}  
			\gamma_{\alpha} (\yslash - \xslash) \Big] \\
			&- A^P_{\mu\nu} (x,y) Tr \Big[ \gamma_{\beta} (\yslash - \xslash) 
			\gamma_{\alpha} \Big] 
			- 2 (\yslash - \xslash) \gamma_{\beta}
			B^P_{\mu\nu} (x,y) \gamma_{\alpha}  
			- 2  \gamma_{\beta} B^P_{\mu\nu} (x,y) \gamma_{\alpha} (\yslash - \xslash) \\
			&+ \dfrac{1}{(y-x)^{6}}
			\Big(
			2 (\yslash - \xslash) 
			Tr \Big[ \gamma_{\beta} (\yslash - \xslash) 
			\gamma_{\alpha} A^5_{\mu\nu} (x,y) \Big]
			+ 2 (\yslash - \xslash) \gamma_{\beta} (\yslash - \xslash) 
			\gamma_{\alpha} A^5_{\mu\nu} (x,y) \\
			&+ 2 A^5_{\mu\nu} (x,y) \gamma_{\beta} (\yslash - \xslash)
			\gamma_{\alpha} (\yslash - \xslash) 
			+  A^5_{\mu\nu} (x,y) 
			Tr \Big[ \gamma_{\beta} (\yslash - \xslash) 
			\gamma_{\alpha} (\yslash - \xslash) \Big] 
			+ 2 (\yslash - \xslash) \gamma_{\beta}
			B^5_{\mu\nu} (x,y) \gamma_{\alpha} (\yslash - \xslash)
			\Big)  \Big\},
		\end{aligned}
	\end{align}
\end{widetext}
where,
\begin{widetext} 
\begin{align}\label{eq:A}
\begin{aligned}
&A^P_{\mu\nu} (x,y) = \dfrac{\yslash}{y^4}\gamma_{\nu} 
\Big[\dfrac{\gamma_{\mu}}{x^4} - \dfrac{4 \xslash x_{\mu}}{x^6}
\Big] - \Big[\dfrac{\gamma_{\mu}}{y^4} - \dfrac{4 \yslash y_{\mu}}{y^6}
\Big] \gamma_{\nu} \dfrac{\xslash}{x^4}, 
\\
&A^3_{\mu\nu} (x,y) =	 \gamma_{\nu} 
\Big[\dfrac{\gamma_{\mu}}{x^4} - \dfrac{4 \xslash x_{\mu}}{x^6}
\Big]
+ \Big[\dfrac{\gamma_{\mu}}{y^4} - \dfrac{4 \yslash y_{\mu}}{y^6}
\Big] \gamma_{\nu},
\\
&A^{G^2}_{\mu\nu} (x,y)	=	\Big[
\dfrac{\yslash \sigma^{\lambda\delta} + \sigma^{\lambda\delta}\yslash}{y^2} \Big]
\gamma_{\nu} \Big[ \Big(
\dfrac{\gamma_{\mu}}{x^2} - \dfrac{2 \xslash x_{\mu}}{x^4}
\Big) \sigma_{\lambda\delta}
+ \sigma_{\lambda\delta}
\Big( \dfrac{\gamma_{\mu}}{x^2} - \dfrac{2 \xslash x_{\mu}}{x^4}
\Big) \Big] 
\\
&\qquad\qquad - \Big[ \Big(
\dfrac{\gamma_{\mu}}{y^2} - \dfrac{2 \yslash y_{\mu}}{y^4}
\Big) \sigma^{\lambda\delta}
+ \sigma^{\lambda\delta}
\Big( \dfrac{\gamma_{\mu}}{y^2} - \dfrac{2 \yslash y_{\mu}}{y^4} \Big)
\Big] \gamma_{\nu}
\Big[
\dfrac{\xslash \sigma_{\lambda\delta} + \sigma_{\lambda\delta}\xslash}{x^2} \Big],	
\\
&A^{PG}_{\mu\nu} (x,y) = \dfrac{\yslash}{y^4}
\gamma_{\nu} \Big[ \Big(
\dfrac{\gamma_{\mu}}{x^2} - \dfrac{2 \xslash x_{\mu}}{x^4}
\Big) \sigma_{\lambda\delta}
+ \sigma_{\lambda\delta}
\Big( \dfrac{\gamma_{\mu}}{x^2} - \dfrac{2 \xslash x_{\mu}}{x^4}
\Big) \Big] 
- \Big[\dfrac{\gamma_{\mu}}{y^4} - \dfrac{4 \yslash y_{\mu}}{y^6}
\Big] \gamma_{\nu}
\Big[
\dfrac{\xslash \sigma_{\lambda\delta} + \sigma_{\lambda\delta}\xslash}{x^2} \Big]
\\
&\qquad\qquad+ \Big[
\dfrac{\yslash \sigma_{\lambda\delta} + \sigma_{\lambda\delta}\yslash}{y^2} \Big]
\gamma_{\nu} \Big[\dfrac{\gamma_{\mu}}{x^4} - \dfrac{4 \xslash x_{\mu}}{x^6}
\Big]
-\Big[
\Big(
\dfrac{\gamma_{\mu}}{y^2} - \dfrac{2 \yslash y_{\mu}}{y^4}
\Big) \sigma_{\lambda\delta}
+ \sigma_{\lambda\delta}
\Big(
\dfrac{\gamma_{\mu}}{y^2} - \dfrac{2 \yslash y_{\mu}}{y^4}
\Big)
\Big] \gamma_{\nu} \dfrac{ \xslash}{x^4},
\\
&A^5_{\mu\nu} (x,y) =
\dfrac{2 \gamma_{\nu} \xslash  y_{\mu}}{x^4}
- y^2 \gamma_{\nu} \Big[\dfrac{\gamma_{\mu}}{x^4} - 
\dfrac{4 \xslash x_{\mu}}{x^6} \Big] 
- \Big[\dfrac{\gamma_{\mu}}{y^4} - \dfrac{4 \yslash y_{\mu}}{y^6}
\Big] \gamma_{\nu} x^2
+\dfrac{2\yslash \gamma_{\nu} x_{\mu}}{y^4},
\\
&B_{\mu\nu} (x,y) = A_{\mu\nu} (y,x).
\end{aligned}
\end{align}
\end{widetext}

\section{Gluon condensation} \label{gluoncondens}

The light quark propagator in Eq.~\eqref{eq:lightquarkProp} includes one gluon strength field tensor $- i {g_s G_{ij}^{\lambda\delta}}[\xslash \sigma_{\lambda\delta} + \sigma_{\lambda\delta} \xslash]/{32 \pi^2 x^2} 
$. A two-gluon condensation can be formed by multiplying these terms together in the presence of vacuum. We simplify such expressions with these notations ~\cite{Barsbay:2022gtu},
\begin{equation}\label{eq:G formula}
G^{\alpha\beta}_{ab} = G^{\alpha\beta}_{A} t^{A}_{ab}, \qquad 
t^{A} = \frac{1}{2} \lambda^{A}, \qquad
G^2 = G^{A}_{\alpha\beta} G^{A}_{\alpha\beta}, \qquad
t^{A}_{ab} t^{A}_{a'b'} = \frac{1}{2}
\bigg(\delta_{ab'}\delta_{a'b} - \frac{1}{3}\delta_{ab}\delta_{a'b'}
\bigg),
\end{equation}
where $a, b = 1, 2, 3$ and $A = 1, 2, ..., 8$ are color indices of the fundamental (quark) and the adjoint (gluon) representations, respectively and $\lambda^{A}$ are Gell-Mann matrices. We consider $\langle0|G^{A}_{\alpha\beta}(x) G^{A'}_{\alpha'\beta'}(0)|0\rangle$ as the gluon condensate and use the first term of the Taylor expansion at $x=0$,
\begin{equation}\label{eq:GG formula}
\langle0|G^{A}_{\alpha\beta}(0) G^{A'}_{\alpha'\beta'}(0)|0\rangle
= \frac{\langle G^2\rangle}{96} \delta^{A A'}
\bigg[g_{\alpha\alpha'} g_{\beta\beta'}
- g_{\alpha\beta'} g_{\alpha'\beta}
\bigg].
\end{equation}
We apply Eqs.~\eqref{eq:G formula} and \eqref{eq:GG formula} to Eqs.~\eqref{eq:gcontractions} and \eqref{eq:4Dquark}.

\begin{acknowledgments}

Z. Dehghan and K. Azizi are thankful to Iran Science Elites Federation (Saramadan) for the partial financial support provided under  Grant No. ISEF/M/401385.

\end{acknowledgments}

\nocite{*}

\bibliographystyle{elsarticle-num}
\bibliography{refs.bib}

\end{document}